\documentclass[12pt]{article}

\usepackage{amssymb}
\usepackage{authblk}
\usepackage{comment}
\usepackage{amsmath,empheq,systeme}
\usepackage[hidelinks]{hyperref}
\usepackage[dvipsnames]{xcolor}
\usepackage{graphicx}
\usepackage{caption}
\usepackage{subcaption}
\usepackage{siunitx}
\DeclareSIUnit\bar{bar}
\usepackage{booktabs}
\usepackage{multirow}
\usepackage{textcomp}
\usepackage{gensymb}
\usepackage[version=4]{mhchem}
\usepackage[numbers,sort&compress]{natbib}
\usepackage[capitalize]{cleveref}
\usepackage{tikz}
\usetikzlibrary{positioning}
\usetikzlibrary{arrows.meta}
\usetikzlibrary{fit}
\usepackage{standalone}

\title{A zero-dimensional global conservation model to determine non-ideal nozzle conditions for gas injections}

\author{%
N. Diepstraten\thanks{Corresponding author. E-mail: \url{n.diepstraten@tue.nl}},\,
L.M.T. Somers,\, 
J.A. van Oijen
}

\affil{\textit{Department of Mechanical Engineering (Power \& Flow Group),\\
Eindhoven University of Technology,\\
Groene Loper 3, 5612 AE Eindhoven,\\
Noord-Brabant, The Netherlands}}

\date{} 

\begin{document}
\maketitle
\vspace{-1.5em}

\begin{abstract}
To reduce computation times of simulations involving gas fueled internal combustion engines (ICEs), a model is developed that determines non-ideal nozzle exit conditions to spare expensive simulations of internal injector flows. 
The model, to which we will refer as the Global Conservation Model (GCM), computes nozzle exit values based on reservoir conditions, and a discharge and momentum coefficient. 
These coefficients can be obtained via common flow bench experiments or a validated numerical simulation setup.
Furthermore, it allows to use real gas thermodynamics which is often needed for ICE applications. The model is validated for three commonly used injector types, injection pressures up to \SI{300}{\bar}, subsonic and choked injections, and at flow bench and engine conditions using numerical simulations. 
If real gas thermodynamics is applied, differences with the simulated nozzle conditions are typically within \SI{1.5}{\percent} uncertainty, which is within the typical range of experimental momentum flow measurements found in literature. 
Differences in mass and momentum flow obtained by the numerical simulations between flow bench and engine conditions are negligible, indicating that flow bench measurements can be applied to engine conditions. 
Once an injector is characterized for mass and momentum flow, applying the GCM can reduce computation times by more than a factor 8, while accurately simulating the spatial and temporal jet development. 
With the reduction in computation time, the GCM reduces costs associated with numerical simulations and accelerates research and design of efficient and low-emission gas fueled ICEs.
\end{abstract}

\section{Introduction}
\label{sec:introduction}
\noindent In 2015, 196 Parties agreed to a legally binding international treaty, The Paris Agreement, at the United Nations Climate Change Conference (COP21) to combat climate change by reducing greenhouse gas (GHG) emissions~\citep{unfccc2015adoption}. 
The main contributor of GHG emissions is carbon dioxide (\ce{CO2}), accounting for around \SI{72}{\percent} of total emissions in 2022~\citep{crippa2021ghg}.
According to the IEA, the transport sector was responsible for \SI{23}{\percent} of the total \ce{CO2} emissions in 2021~\citep{iea_data}. 
The internal combustion engine (ICE) is the predominant technology in this sector, and it is expected to remain of great significance in the foreseeable future~\citep{onorati2022role,reitz2020ijer,kalghatgi2015developments}.
Various initiatives are taken across the globe to decarbonize the transport sector~\citep{greendeal,muratori2023us,unctad2023chinapolicy}. In Europe, the European climate law has adopted the package called Fit For 55~\citep{fit_for_55}. This makes the goal of reducing European Union's emissions by at least \SI{55}{\percent} by 2030 a legal obligation. Regarding the transport sector, one of the proposals to achieve this is to progressively replace fossil gas in the EU with renewable and low-carbon gases, such as methane and hydrogen.
It is expected that the medium- and heavy-duty transport sector, which is now for around \SI{95}{\percent} fueled by diesel, will consist of a mix of hydrogen, natural gas, bio-methane and electricity \citep{bp2023energyoutlook}.
The employment of gaseous fuels in ICEs is therefore a highly relevant research topic.

A drawback of gaseous fuels compared to liquid fuels is their relatively low volumetric energy density. When employing port-fuel injection (PFI), this results in a penalty in volumetric efficiency, which can be large especially for hydrogen. Direct-injection (DI) technologies can mitigate this. 
To develop efficient and low-emission gas fueled DI engines, accurate and computationally efficient numerical models are needed to reduce expensive prototyping and to gain detailed insights in in-cylinder processes which cannot be obtained via experiments.

There are various DI combustion concepts, which can be divided into low-pressure DI (LPDI) and high-pressure DI (HPDI). In case of LPDI, the gaseous fuel injection happens early in the compression stroke, while the in-cylinder pressure is relatively low (\textless40 bar). Injection pressures typically vary up to \SI{40}{\bar}~\citep{mohamed2024experimental}. In case of HPDI, the fuel is injected late in the compression stroke and requires a high injection pressure (typically between 200 and \SI{350}{\bar})~\citep{mctaggart2015direct,fink2018influence}. 
Besides the different conditions, also various injector designs exist. For LPDI, often outwardly opening poppet-valve type injectors are employed because such a design seals the valve by the in-cylinder pressure and can facilitate a high mass flow rate at relatively low fuel pressures \citep{deshmukh2018characterization,SANKESH2018188}. 
However, for HPDI a multi-hole design similar to diesel injectors is used thanks to its favorable fuel distribution~\citep{mctaggart2015direct}.

Accurate simulation of the injection process is paramount for researching DI concepts, as it determines the spatial distribution of the fuel and flow field at the time ignition occurs. 
Experimental studies have reported discharge coefficients in the range of 0.43--0.9 for cylindrical nozzles \citep{peters2024characterizing,yip2020visualization,tsujimura2003study}, which indicates that the injector flow can be far from ideal such that simple relations to describe a thermodynamic ideal, sonic injection do not suffice.
Therefore, it is important to account for non-ideal flow effects caused by the specific injector geometry. For this reason, advanced approaches are required to accurately determine nozzle exit conditions as function of reservoir conditions. One option to achieve this is to incorporate (a part of) the internal geometry of the injector in the simulation domain. 
Such studies have been performed for cylindrical nozzles~\citep{maes2023hydrogen,banholzer2018numerical} and outwardly opening poppet-valve type injectors~\citep{keskinen2016mixture,leick2023analysis,baratta2009multi,deshmukh2018characterization}. However, this approach is very expensive as it requires high resolution in space and time to accurately simulate the high velocity flow in the small, complicated injector geometry. Besides, for most engine research, the area of interest is the flow in the combustion chamber and not in the injector. Therefore, it is preferred to simulate the combustion chamber domain only.

To reduce computation times, it is convenient to separate the internal injector flow from the combustion chamber and compute the nozzle exit conditions using a computationally cheaper yet reasonably accurate model \citep{ouellette1998numerical,deshmukh2020quasi}.
There are several studies found in literature reporting methods to compute nozzle conditions in such a way. These studies range from three-dimensional (3D) methods, which use computational fluid dynamics (CFD) simulations to derive nozzle exit conditions, to zero-dimensional (0D) methods, which use thermodynamic relations to determine nozzle exit conditions based on reservoir conditions.
The 3D method has been applied to cylindrical nozzles \citep{muller2013determination,banholzer2018numerical,schmitt2015multiple} as well as for outwardly opening poppet-valve type of injectors \citep{deshmukh2018numerical, baratta2011multidimensional}. After the quasi-steady-state injection conditions are computed, they are mapped on the in-cylinder simulation domain near the nozzle exit. This method allows to retain possible spatial variations (albeit time-averaged), but is still computationally expensive, especially when this procedure needs to be repeated for every injection condition.

Computation times can be further reduced by computing the internal injector flow in 1D. 
One option encountered in literature is assuming a Fanno flow in the injector \citep{mancaruso2024high}, thereby assuming an adiabatic flow. In this study, an average Fanning factor to calculate friction stress was assumed and variations in cross-sectional flow profile were not accounted for. Despite the assumptions, the 1D model was in reasonable agreement with the 3D CFD simulations, showing deviations in nozzle conditions up to \SI{11}{\percent}.
In another method, introduced by~\citet{deshmukh2020quasi}, the internal geometry of an outwardly opening poppet-valve type of injector is approximated using a 1D profile with variable cross-section. The internal injector flow is computed by 1D compressible Euler equations and dynamically coupled to the 3D CFD model through source terms in the governing equations. 
A drawback of 1D and 0D methods can be their inherent limitation to compute mean flow properties only, and thereby loosing the capability to compute spatially varying cross-sectional flow fields. However, it has been shown for methane injections emanating from a poppet-valve type injector that the boundary layer in the nozzle is very thin and has a minor impact on the flow, except at the exit where the flow starts separating \citep{deshmukh2020quasi,deshmukh2018characterization}. 

The above mentioned methods use numeric models to determine the internal injector flow. Hence, there is a missing link with experimental data such that the methods purely rely on the accuracy of the used numerical model.
Furthermore, these methods require the internal geometry of the injector as a model input. This is not obvious as it is often proprietary information. Deriving the injector geometry, for example using a Computer Tomography (CT) scan, requires specialized and expensive equipment and is therefore not that common. An accurate 0D approach, which does not need detailed information of the internal geometry, can alleviate this.

Studies exploring 0D methods dealing with non-ideal flows are scarce. In \citet{ouellette2000turbulent}, exit conditions for cylindrical nozzles are calculated by deriving a nozzle momentum rate based on the experimentally measured jet penetration. The nozzle exit velocity and temperature are assumed to be equal to the ideal values, which might not be valid. Subsequently, the nozzle pressure is adjusted to match the measured momentum rate. A drawback of this method is that it does not constrain the mass flow rate, which is problematic for engine simulations since it directly relates to the engine's power output.
A similar model was adopted by \citet{kim2007supersonic} to describe the injection process of a poppet-valve type injector.  
The annular nozzle exit plane of the injector was approximated by a converging-diverging cylindrical nozzle, and it was shown to approximate the downstream part of the jet. However, due to simplification of the nozzle exit geometry, the model is incapable to accurately simulate the near nozzle flow field. 
Lastly, the experimental study of \citet{yip2020visualization} estimates the nozzle exit conditions of a cylindrical nozzle by simplifying the flow through the nozzle with a certain discharge coefficient as a converging-diverging flow through a contraction. Although the model considers a choked flow through a converging-diverging geometry which should result in a supersonic nozzle exit flow, the computed nozzle exit conditions are subsonic. 
Furthermore, Schlieren measurements are required to determine the Mach disk location, which requires expensive equipment. 
Notably, none of these studies have been proven to be universally valid for both cylindrical nozzles and poppet-valve type injectors. 

\begin{figure}
    \centering
    \includegraphics[width=0.8\linewidth]{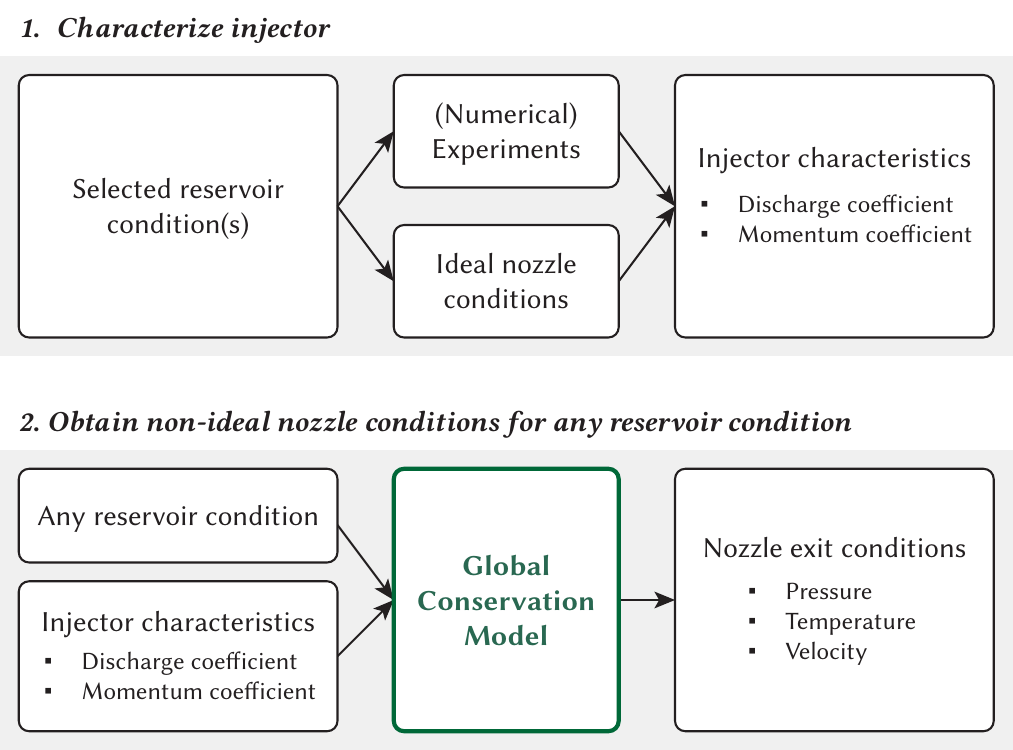}
    \caption{Schematic overview of the process to determine non-ideal nozzle exit conditions.}
    \label{fig:flow_diagram}
\end{figure}

This study proposes a 0D approach to determine non-ideal nozzle exit conditions. It is based on the presumption that the injector's non-ideal flow behavior can be characterized by a discharge coefficient, $C_\mathrm{D}$, and a momentum coefficient, $C_\mathrm{M}$, thereby fixing both mass and momentum injection flow rate.
These coefficients can be obtained through flow bench measurements which are commonly performed by manufacturers. Alternatively, they can be obtained with a validated CFD simulation setup. 
This model, to which we will refer as the global conservation model (GCM), solves a mathematical system of equations that describes the conservation of flow properties between the reservoir and the nozzle exit.
Real gas thermodynamics is included by use of the thermodynamic library CoolProp~\citep{coolprop}.
Figure~\ref{fig:flow_diagram} schematically illustrates the process to determine non-ideal nozzle exit conditions.

Since nozzle exit conditions cannot be affected by the downstream conditions in case of choked flow, the mass and momentum flow rates can be measured experimentally without perturbing the internal nozzle flow, and with relatively cheap measurement techniques. The mass flow rate is typically measured in a constant volume chamber by measuring the pressure rise resulting from the injection process, such as in the study of \citet{peters2024characterizing}, where mass flow rates of high-pressure hydrogen injections were measured. The momentum flow rate can be measured by installing a pressure sensor in front of the nozzle exit. Although most existing literature regarding mass flux measurements concerns liquid fuels, the validity of this method is shown for gaseous fuels as well, such as methane \citep{faghani2015application} and hydrogen \citep{postrioti2024experimental}.
With the use of injector characteristics to relate ideal to non-ideal injection properties, the proposed model shows similarities with the widely used method for liquid fuels, which assumes incompressible flow. This method was introduced by \citet{naber1996effects} and has been adopted successfully in many subsequent studies \citep{payri2008diesel,musculus2013conceptual,postrioti2012experimental,payri2005using,payri2009study}. 

In Section~\ref{sec:0d_model}, the GCM to determine non-ideal nozzle exit conditions is explained. The ideal nozzle exit conditions are briefly discussed first. Subsequently, the GCM is explained for various model assumptions. 
Section~\ref{sec:sim_setup} discusses the CFD setups used to validate the 0D model. The validation is demonstrated for a variety of injections, first by various injections through a single-hole injector at flow bench conditions in Section~\ref{sec:validation}, followed by injections using a poppet-valve type and a multi-hole injector at engine-like conditions in Section~\ref{sec:case_studies}.
In Section~\ref{sec:adequacy}, the adequacy of different model assumptions are studied, followed by a sensitivity study of the discharge and momentum coefficient on nozzle exit conditions in Section~\ref{sec:sensitivity}. Lastly, the effects of the simplification of the injector flow on the resulting jet simulation and on the associated computational costs are studied in Section~\ref{sec:effect}.

\section{Zero-dimensional nozzle conditions}
\label{sec:0d_model}
\noindent To describe the nozzle exit conditions, at least three properties need to be known: one physical property, such as velocity, and two state variables to fix the thermodynamic state. For the latter, generally pressure and temperature are used, but any combination suffices.
The goal is to determine nozzle exit conditions based on reservoir conditions and the nozzle mass and momentum flow rates. Similar to the method for liquid fuels \citep{naber1996effects}, the non-ideal nozzle mass and momentum flow rates for an arbitrary injection condition can be determined with the ideal flow rates and injector characteristics $C_\mathrm{D}$ and $C_\mathrm{M}$.
First, ideal nozzle conditions are briefly discussed, as they are needed to estimate $C_\mathrm{D}$ and $C_\mathrm{M}$. Subsequently, the method to determine the non-ideal injection conditions is discussed.

\subsection{Ideal nozzle conditions}
\noindent Ideal nozzle conditions are found in the absence of any non-ideal effect, such as friction or heat transfer.
For a gaseous injection, these conditions can be computed by considering an isentropic, isenthalpic expansion between the reservoir (where $U=0$) and the nozzle exit. If the upstream-to-downstream pressure ratio is large enough, the flow will choke at the throat, i.e.\@ at the smallest cross-sectional flow area. At the throat, the flow will reach the sonic condition, meaning that the velocity $U$ equals the local speed of sound. For the ideal nozzle conditions it is assumed that the throat is located at the nozzle exit. Consequently, these conditions can be computed by solving the following system of equations:
\begin{subequations}\label{eq:soe_th}
    \begin{empheq}{align}
        s(\rho, T) &= s(\rho_\mathrm{res}, T_\mathrm{res})\\
        h(\rho, T) + \frac{1}{2}U^2 &= h(\rho_\mathrm{res}, T_\mathrm{res})\\
        U &= \sqrt{\frac{\partial P}{\partial \rho}\Big|_s}\\
        P &= f_\mathrm{EOS}(\rho, T)
    \end{empheq}
\end{subequations}
where $s$ is the entropy, $\rho$ the density, $T$ the temperature, $h$ the enthalpy and $P$ the pressure. Subscripts res refers to the reservoir. It is good to note that properties without subscript refer to the (static) nozzle exit properties. Lastly, $f_\mathrm{EOS}$ indicates an equation of state (EOS) providing pressure as function of density and temperature. Equation~\ref{eq:soe_th} describe conservation of entropy and total energy, while demanding the nozzle exit velocity to be sonic via the definition of the speed of sound. The EOS relates the thermodynamic state variables, which is needed to close the mathematical problem.

Other relevant injection properties, such as the ideal mass flow rate and momentum flow rate, can be determined with these equations, respectively: 
\begin{equation}
\label{eq:mdot_pg}
    \Dot{m}_\mathrm{id} = \left(\rho U\right)_\mathrm{id}A ,
\end{equation}
\begin{equation}
\label{eq:Mdot_pg}
    \Dot{M}_\mathrm{id} = \left(\rho U^{2}\right)_\mathrm{id} A  ,
\end{equation}
where $A$ is the nozzle exit surface area and subscript id refers to an ideal situation. 
Alternatively, explicit equations can be derived to relate the reservoir conditions to the nozzle exit conditions when perfect gas behavior is assumed. For an elaborate explanation to derive the nozzle exit conditions with other gas laws, refer to our previous work~\citep{DIEPSTRATEN202422}. For a perfect gas, the nozzle pressure, density and velocity can be expressed as:
\begin{equation}
    P_\mathrm{id,pg} = P_\mathrm{res} \left( \frac{2}{\gamma+1} \right)^{\frac{\gamma}{\gamma-1}} ,
\end{equation}
\begin{equation}
    \rho_\mathrm{id,pg} = \rho_\mathrm{res} \left( \frac{2}{\gamma+1} \right)^{\frac{1}{\gamma-1}} ,
\end{equation}
\begin{equation}
    U_\mathrm{id,pg} = \sqrt{\gamma \left(\frac{P}{\rho}\right)_\mathrm{id,pg}} ,
\end{equation}
where $\gamma=c_p/c_v$ is the ratio of specific heats. Subscript pg denotes that it concerns a perfect gas.

\subsection{Non-ideal nozzle conditions \label{sec:act_noz_conds}}
\noindent In previous section, we computed sonic injection properties assuming an isentropic, isenthalpic process between the reservoir and the nozzle exit. In other words, we equated the reservoir conditions to the stagnation properties of the nozzle exit, i.e.\@ the thermodynamic state if we would decelerate the gas to rest while conserving entropy and total energy. 

In reality, the nozzle exit velocity may not be sonic and total energy and entropy will not be conserved between the reservoir and the nozzle exit due to non-ideal flow effects, such as friction. It is therefore inappropriate to use the reservoir pressure and temperature as stagnation conditions at the nozzle exit. The non-ideal flow behavior of the injector will result in a mass flow rate lower than the theoretical value (\ref{eq:mdot_pg}) and may result in a momentum flow rate lower or higher than its theoretical value (\ref{eq:Mdot_pg}). 
To characterize the deviation from ideal flow behavior of an injector, a discharge coefficient, $C_\mathrm{D}$, and a momentum coefficient, $C_\mathrm{M}$ are used.
These coefficients can be constant or a function of the reservoir properties. They are defined as:
\begin{equation}\label{eq:cd}
    C_\mathrm{D} = \frac{\dot{m}}{\dot{m}_\mathrm{id}} ,
\end{equation}
\begin{equation}\label{eq:cm}
    C_\mathrm{M} = \frac{\dot{M}}{\dot{M}_\mathrm{id}} .
\end{equation}
The definitions for the mass and axial momentum flow rate can be described by: 
\begin{equation}
    \label{eq:mdot_def}
    \dot{m} = \int_A \rho \left( \vec{\boldsymbol{u}} \cdot \vec{\boldsymbol{n}} \right) dA ,
\end{equation}
\begin{equation}
    \label{eq:Mdot_def}
    \dot{M} = \int_A \rho \left( \vec{\boldsymbol{u}} \cdot \vec{\boldsymbol{n}} \right)^2 dA ,
\end{equation}
where $\rho$ is the density and $\left( \vec{\boldsymbol{u}} \cdot \vec{\boldsymbol{n}} \right)$ indicates the velocity normal to $A$ if it is assumed that it is the axial component. If the density and velocity are assumed to be uniform across $A$, which is necessary for a 0D approach, these expressions reduce to:
\begin{equation}
    \label{eq:mdot0D}
    \dot{m} = \rho U A ,
\end{equation}
\begin{equation}
\label{eq:Mdot0D}
    \dot{M} = \rho U^2 A ,
\end{equation}
where $U$ indicates the mean normal velocity. 
Notably, if $A$, $C_\mathrm{D}$ and $C_\mathrm{M}$ are known, the density and velocity can be obtained for certain reservoir conditions by combining Eqs.~\ref{eq:cd}, \ref{eq:cm}, \ref{eq:mdot0D} and \ref{eq:Mdot0D}:
\begin{equation} \label{eq:U_expl}
    U =\frac{\dot{M}}{\dot{m}}= \frac{C_\mathrm{M}\dot{M}_\mathrm{id}}{C_\mathrm{D}\dot{m}_\mathrm{id}} = \frac{C_\mathrm{M}}{C_\mathrm{D}}U_\mathrm{id}
\end{equation}
\begin{equation} \label{eq:rho_expl}
    \rho =\frac{\dot{m}^2}{\dot{M}A}= \frac{\left(C_\mathrm{D}\dot{m}_\mathrm{id}\right)^2}{C_\mathrm{M}\dot{M}_\mathrm{id}A} = \frac{C_\mathrm{D}^2}{C_\mathrm{M}}\rho_\mathrm{id}
\end{equation}
As mentioned at the beginning of this chapter, one physical property and two thermodynamic state variables are needed to fully describe the nozzle exit conditions. Since the density and velocity are known, only one more state variable is needed, e.g.\@ the temperature or pressure at the nozzle exit.
Unless the second state variable is known by a measurement, we need to invoke an equation that relates this variable to a known variable. 
For example, one can assume that the pressure at the nozzle exit is the same as the downstream pressure. Although relating a nozzle exit property to a downstream condition may be adequate for subsonic flows, it is not for choked injections~\citep{shapiro1953dynamics}. For such cases, it is more appropriate to make an assumption about the process between the reservoir and the nozzle exit.
This can be any relation as long as it involves at least one state variable which is not the density. 
To compute ideal injection conditions, the process between the reservoir and the nozzle exit is assumed to be both isenthalpic and isentropic. In the following, we will first discuss the approaches associated with either an isenthalpic or isentropic process. As third option, we assume neither of these processes, but seek for an alternative assumption.

One option is to assume an isenthalpic process between the reservoir and the nozzle exit. With this assumption, total energy, i.e.\@ the sum of the internal enthalpy and kinetic energy, is conserved. The resulting system of equations then reads: 
\begin{subequations}
\label{eq:soe_isenthalpic}
    \begin{empheq}{align}
        \rho A U   &= C_\mathrm{D} \dot{m}_\mathrm{id} \\ 
        \rho A U^2 &= C_\mathrm{M} \dot{M}_\mathrm{id} \\
        h(\rho_\mathrm{res},T_\mathrm{res}) &= h(\rho,T)+\tfrac{1}{2}U^2 \label{eq:soe_enthalpy}
    \end{empheq}
\end{subequations}
Note that in this system, entropy does not need to be conserved between the reservoir and the nozzle exit. 
Considering the first law of thermodynamics and the fact that there is no work done by the fluid, it is obvious that these equations describe an adiabatic flow in the injector. Since the second law of thermodynamics for an irreversible and actually possible process states that $\mathrm{d}s > \delta Q/T$, where $Q$ is heat, the entropy can only increase between the reservoir and nozzle exit due to irreversible processes like friction and turbulence.

\begin{figure}
    \centering
    \includegraphics[width=0.6\linewidth]{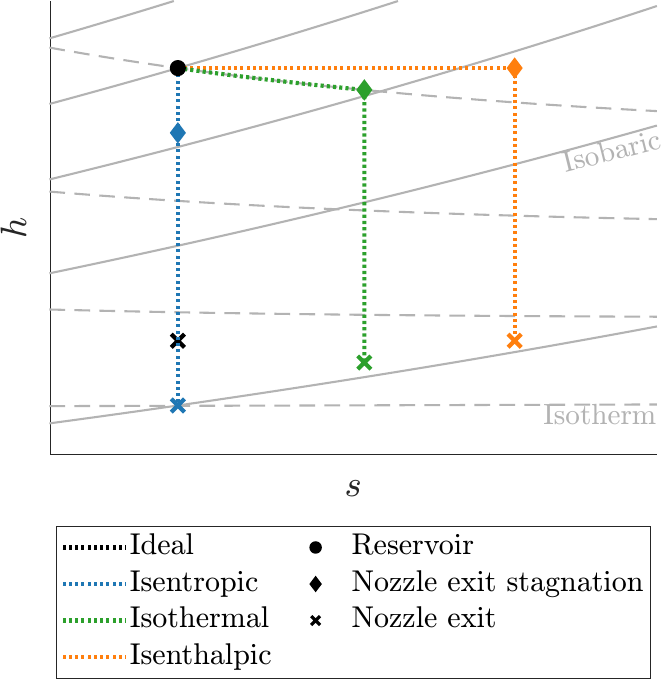}
    \caption{Enthalpy vs. entropy plots for the discussed 0D model approaches. Note that the enthalpy difference between the nozzle exit state ($\times$) and its stagnation state ($\blacklozenge$) are equal for all approaches due to the fixed kinetic part ($U^2/2$).}
    \label{fig:hs}
\end{figure}

A second option would be to assume that entropy is conserved between the reservoir and the nozzle exit, but not enthalpy. With this assumption, the non-ideal flow effects are purely translated to a decrease in total energy.
As mentioned above, the entropy of a fluid increases due to irreversible processes. It can only decrease if heat is removed from the system. Therefore, an isentropic process occurs if the internal heat generation and heat transfer from the fluid to the injector walls are in balance.
This approach may not be accurate at engine conditions, since the temperature of the injector walls will be higher than the temperature of the fluid due to the expansion, but it might be acceptable at test bench conditions where heat transfer is small.
In that case, the system of equations reads:
\begin{subequations}
\label{eq:soe_isentropic}
    \begin{empheq}{align}
        \rho A U   &= C_\mathrm{D} \dot{m}_\mathrm{id} \\ 
        \rho A U^2 &= C_\mathrm{M} \dot{M}_\mathrm{id} \\
        s(\rho_\mathrm{res},T_\mathrm{res}) &= s(\rho,T) 
    \end{empheq}
\end{subequations}
A third option is to neither assume conservation of enthalpy or entropy. In that case, another property needs to be conserved. For example, we can assume that the stagnation temperature at the nozzle exit is equal to the reservoir temperature. In other words, if the nozzle exit flow is decelerated to standstill by means of a hypothetical isenthalpic, isentropic process, the temperature is equal the reservoir temperature. The process between the reservoir and the stagnation state at the nozzle exit happens at a constant stagnation temperature, which is why it will be referred to as isothermal.
In this approach, both enthalpy and entropy relations need to be considered in order to apply a stagnation property, such that the system of equations reads:
\begin{subequations}
\label{eq:soe_isothermal}
    \begin{empheq}{align}
        \rho A U   &= C_\mathrm{D} \dot{m}_\mathrm{id} \\ 
        \rho A U^2 &= C_\mathrm{M} \dot{M}_\mathrm{id} \\
        s(\rho_\mathrm{stag},T_\mathrm{res}) &= s(\rho,T) \\
        h(\rho_\mathrm{stag},T_\mathrm{res}) &= h(\rho,T)+\tfrac{1}{2}U^2 
    \end{empheq}
\end{subequations}
where $\rho_\mathrm{stag}$ refers to the stagnation density, which is not equal to the reservoir density due to non-ideal flow effects.
Figure~\ref{fig:hs} shows the three discussed model approaches visualized in an enthalpy versus entropy diagram, where the model is characterized by the process between the reservoir and the stagnation state at the nozzle exit.
Isothermal and isobaric lines are added to indicate the processes relative to pressure and temperature. 
Note that nozzle exit velocity is determined by the mass and momentum flow rates (Eq.~\ref{eq:U_expl}). 
As a result, the enthalpy difference between the nozzle exit state ($\times$) and its stagnation state ($\blacklozenge$), which is the kinetic part ($U^2/2$), are equal for all approaches.

It is also possible to relax the isothermal assumption by relating the nozzle temperature to the reservoir and/or injector wall temperature. This modeled heat transfer could be interesting, if dedicated measurements are performed, such as the ones discussed in literature for diesel sprays~\citep{meijer2012engine,payri2012fuel}. This could be realized by adopting Eq.~\ref{eq:soe_isothermal} and replacing $T_\mathrm{res}$ by:
\begin{equation}
    T_\mathrm{stag} = f(T_\mathrm{res},T_\mathrm{wall})
\end{equation}
where $T_\mathrm{wall}$ is the temperature of the injector wall. In this study, the isothermal assumption of Eq.~\ref{eq:soe_isothermal} is applied, unless mentioned otherwise.

Note that the above system of equations allows for the use of real gas thermodynamics, which was shown to be required for high-pressure injections of hydrogen \citep{DIEPSTRATEN202422} and methane \citep{ouellette1996direct}. 
If the gas can be assumed to behave ideal and to have constant specific heats, i.e.\@ a perfect gas, explicit equations can be derived. For the isenthalpic assumption, the nozzle temperature can be found can be calculated by substituting the enthalpy of a perfect gas, $h=c_pT$, into Eq.~\ref{eq:soe_enthalpy}. After rewriting, the nozzle temperature is derived:
\begin{equation}
\label{eq:expl_T_pg_isenth}
    T = T_\mathrm{res} - \frac{\gamma-1}{2} \frac{U^2}{\gamma R} .
\end{equation}
Note that the enthalpy for a perfect gas depends on temperature only. For this reason, the isothermal and isenthalpic approach are identical. For the isentropic assumption, the nozzle temperature can be expressed by the isentropic relation:
\begin{equation} 
\label{eq:expl_T_pg_isentr}
    T = T_\mathrm{res}\left(\frac{\rho}{\rho_\mathrm{res}}\right)^{\gamma-1} .
\end{equation} 
Together with the equations for the nozzle exit velocity (\ref{eq:U_expl}) and density (\ref{eq:rho_expl}), the nozzle exit flow is fully described by explicit equations.

In the following chapters, the accuracy of the 0D model with the isothermal assumption is demonstrated using numerical simulations of injector flows employing various injector designs. Details of the simulation setups are described in the next chapter.
After that, the nozzle exit conditions will be computed with the 0D model using real gas thermodynamics (Eq.~\ref{eq:soe_isothermal}) as well as using a perfect gas assumption (Eqs.~\ref{eq:U_expl}, \ref{eq:rho_expl}, and \ref{eq:expl_T_pg_isenth}) to study the validity of the latter.

\section{Simulation setup}
\label{sec:sim_setup}
\subsection{Computational domains\label{sec:computational_domains}}
\noindent The goal of the CFD simulations is to compute the internal injector flow of three common injector types and to derive nozzle exit conditions to validate the 0D model. 
Accurately determining area-averaged nozzle exit conditions is important in this study. For this reason, a so-called interface boundary is implemented at each nozzle exit plane. This is a completely permeable boundary to which the flow fields are interpolated. Subsequently, area-average values are calculated on this plane.

\begin{figure}
    \centering
    \includegraphics[width=\linewidth]{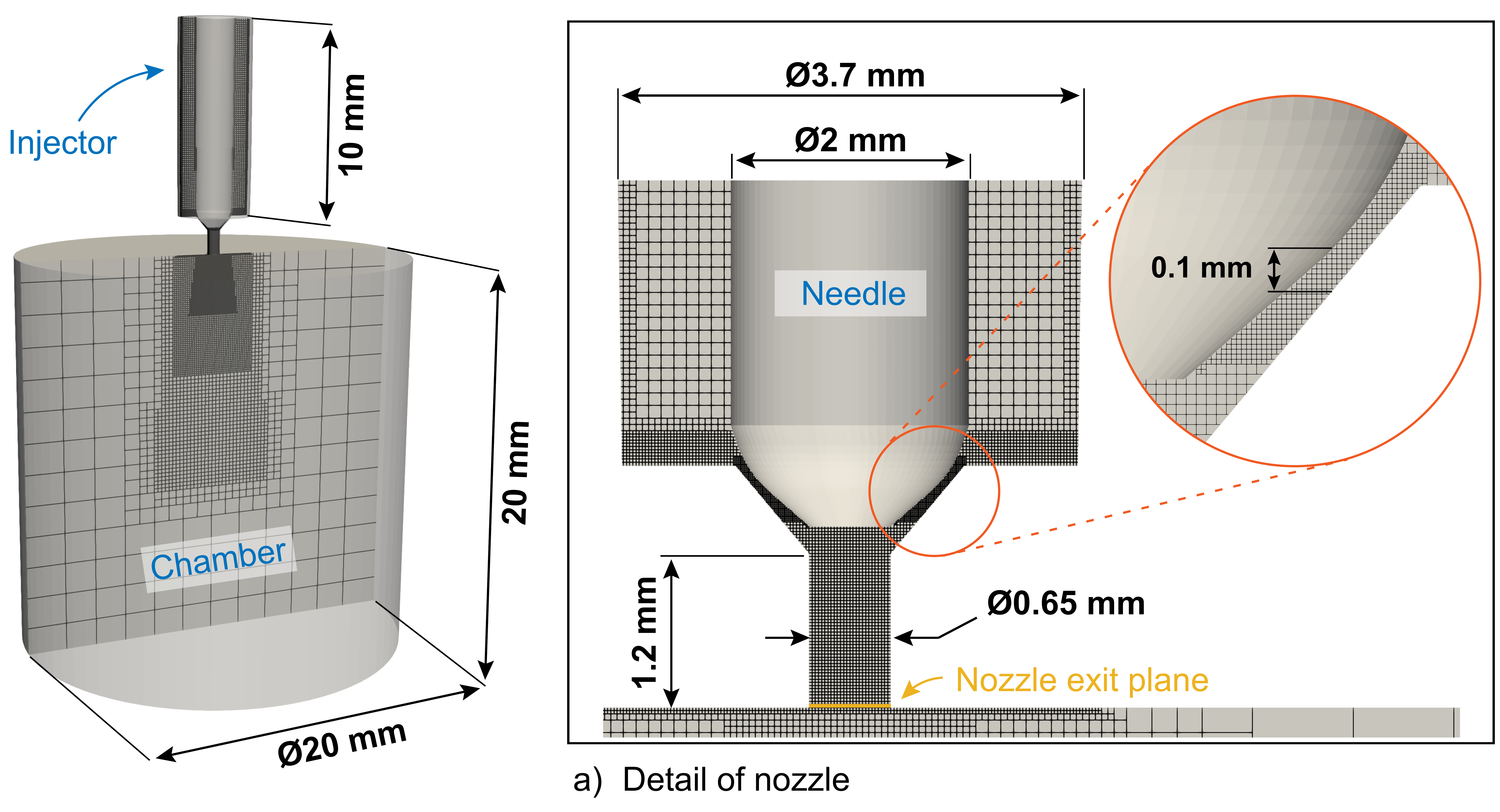}
    \caption{Computational domain of the single-hole injector including some main dimensions.}
    \label{fig:sh_domain}
\end{figure}

The computational domain of the single-hole injector simulations consists of a $360\degree$ geometry of the internal injector volume attached to a small cylindrical chamber. In Fig.~\ref{fig:sh_domain}, the domain is shown including some relevant dimensions. The geometry is shown with a fully retracted needle. As depicted in the detail on the right, the needle lift is set to \SI{0.1}{\milli\m}, which is large enough to locate the throat at the nozzle exit and small enough to cause a discharge coefficient considerably lower than 1. Notably, the geometrical throat would be located at the needle seat for lifts lower than \SI{80}{\milli\m}.

\begin{figure}
    \centering
    \includegraphics[width=\linewidth]{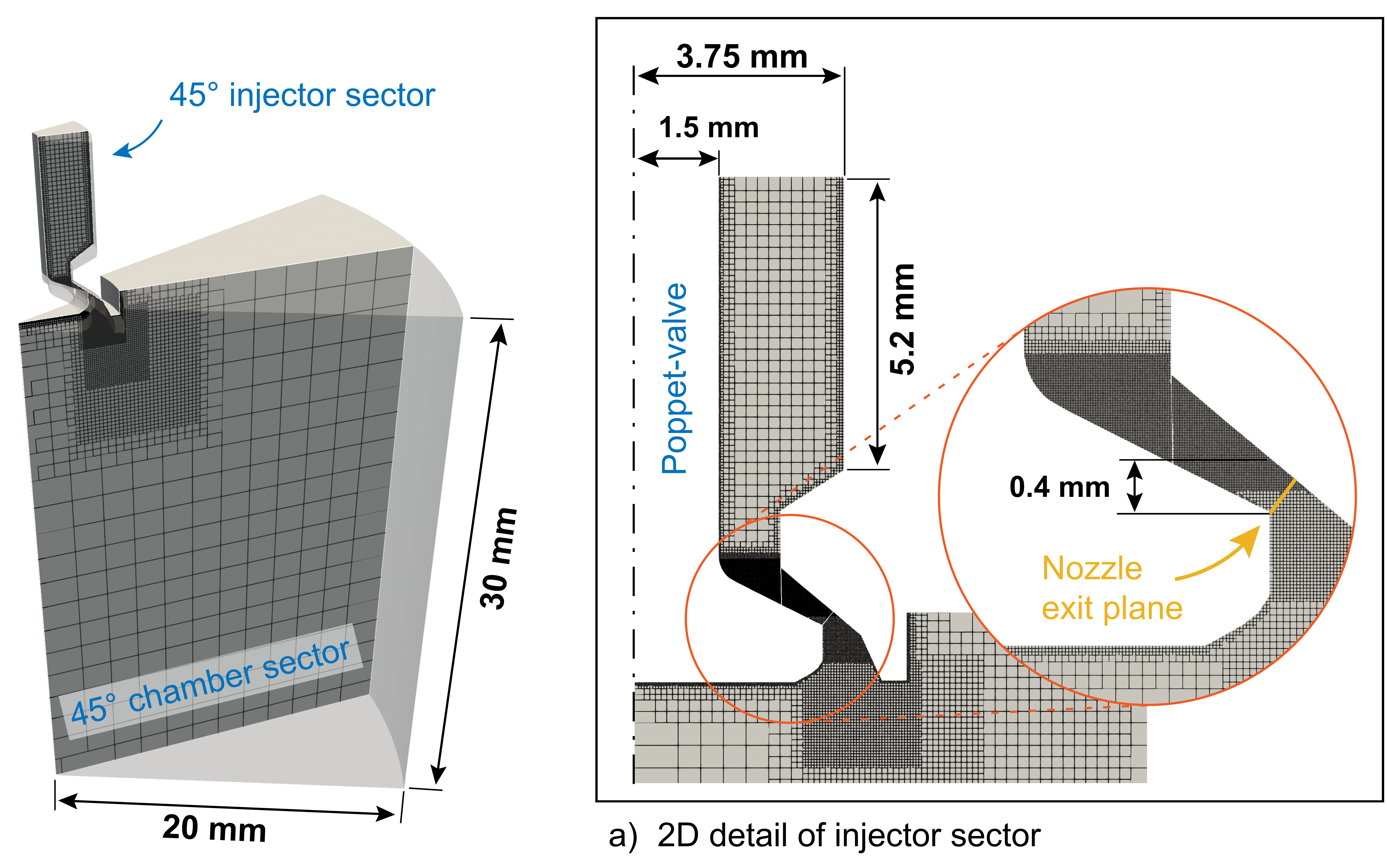}
    \caption{Computational domain of the poppet-valve type injector including some main dimensions.}
    \label{fig:oo_domain}
\end{figure}

Similar to the single-hole injector, the computational domain of the poppet-valve type injector consists of the internal injector geometry attached to a cylindrical chamber. However, only a $45\degree$ sector is simulated to reduce computation times. An overview of the computational domain is provided in Fig.~\ref{fig:oo_domain}. As depicted in the detail on the right, the needle lift is set to \SI{0.4}{\milli\m}. In contrast to the other two geometries, the poppet-valve type injector has an annular nozzle exit plane, which is indicated in yellow in the 2D detail. 

\begin{figure}
    \centering
    \includegraphics[width=\linewidth]{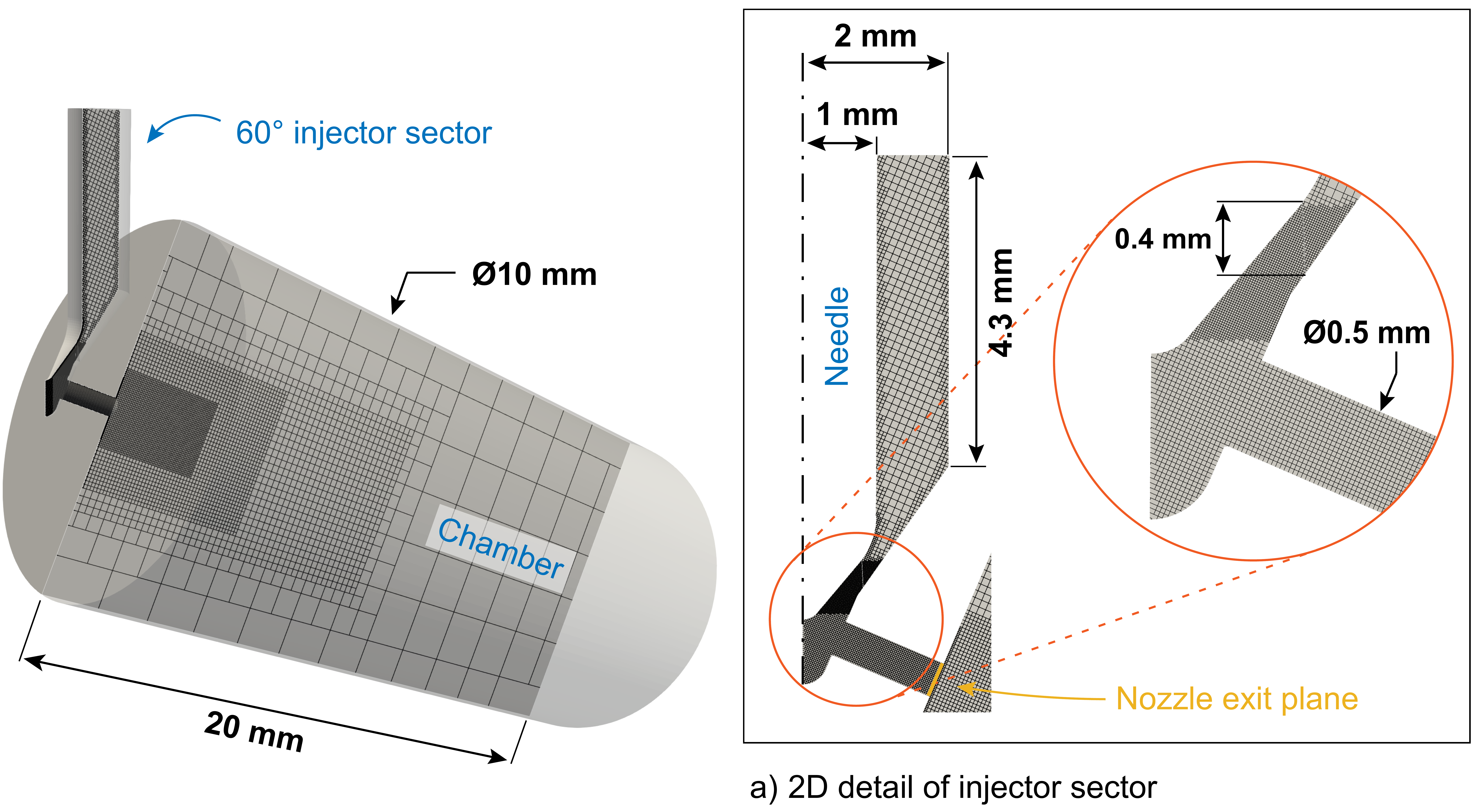}
    \caption{Computational domain of the multi-hole injector including some main dimensions.}
    \label{fig:mh_domain}
\end{figure}

The multi-hole injector domain consists of a $60\degree$ sector of the injector with one hole such that the complete injector contains 6 nozzle holes. A small cylindrical chamber is attached to the nozzle exit, which is visible in Fig.~\ref{fig:mh_domain}. The needle lift is set to \SI{0.4}{\milli\m}, such that the flow chokes at the nozzle. Notice that the needle lift is considerably larger than for the single-hole injector, because the flow along the needle needs to facilitate injections emanating from 6 nozzle holes.

The base grid size of all simulations is set to \SI{1.6}{\milli\m}. On top of that, the mesh is refined locally in the injector and in the vicinity of the nozzle exit. 
The smallest cells are of \SI{12.5}{\micro\m} and can be found in the needle seat area, which are highlighted in Figs.~\ref{fig:sh_domain} to \ref{fig:mh_domain}.
Furthermore, the mesh in the chamber is refined based on the local velocity and temperature with adaptive mesh refinement (AMR) using a so-called sub-grid scale (SGS) criterion~\citep{bedford1993conjunctive,pomraning2000development}. This is a gradient-based refinement method, which is set to reduce the cell size to \SI{0.4}{\milli\m} and \SI{0.2}{\milli\m} for velocity and temperature, respectively.

\subsection{Governing equations}
\noindent The compressible Reynolds-Averaged Navier-Stokes (RANS) equations are solved using the commercial software package CONVERGE~\citep{converge_manual}. 
Regarding the energy transport equation, it is chosen to solve for total energy. Turbulence is modeled with the Renormalization Group (RNG) $k$-$\varepsilon$ model developed by~\citet{yakhot1992development}, where standard values for the model constants are used. This turbulence model is often applied when simulating jets, and has been validated for injections of methane~\citep{papageorgakis1998optimizing} and for high-pressure hydrogen jets into air~\citep{babayev2021computational}.
The thermodynamic fluid properties are obtained from Helmholtz-energy-explicit-type formulations~\citep{coolprop}, which uncertainties approach the level of the underlying experimental data. The fluid properties are tabulated a priori as function of temperature and pressure using open-source library CoolProp. Hence, no equation of state is solved within the CFD code. 
The turbulent Schmidt and Prandtl number are both set to 0.7.

\subsection{Initial and boundary conditions\label{sec:ini_and_bcs}}
\noindent Initially, flow between the injector and the chamber is prohibited due to the needle sealing the two regions. The domain upstream of the needle seat is initialized at reservoir conditions, while the domain downstream (consisting of the nozzle and chamber) is initialized at chamber conditions. The upstream boundary of the injectors is defined as a subsonic inflow boundary and constrained by the reservoir pressure as total pressure. The side and downstream boundaries of the chambers are defined as outflows and constrained by the chamber pressure as total pressure. 
The corresponding values are provided in the chapters discussing the simulation results.
The boundaries of the poppet-valve type and multi-hole injector located at the side planes of the sector are defined as symmetry boundaries. 
The remaining boundaries are modeled as walls with a law-of-the-wall using the standard wall function. Wall heat transfer is modeled using the O'Rourke and Amsden model \citep{amsden1999block}.

The injection is realized by retracting the needle with a linear displacement over time. 
After the needle is fully retracted, the injection is continued until steady nozzle exit conditions are reached. The total simulated time is \SI{0.5}{\milli\s}.

\subsection{Numerics}
\noindent The RANS equations are discretized in space using a blended scheme, which switches between second-order central differencing scheme and a first-order upwind scheme based on the local monotonicity of the flow. To solve the discretized RANS equations, the density-based Pressure-Implicit Splitting of Operators (PISO) algorithm of \citet{issa1986solution} is used. Time-integration is realized by a first-order implicit Euler approach. The time step varies during the simulation based on the Courant-Friedrich-Lewis (CFL) number. The convective CFL limit is kept below 0.5.

\section{Zero-dimensional model validation\label{sec:validation}}
\noindent In this section, the validity of the isothermal 0D model will be demonstrated for a single-hole injector design. Later in this study, in Section~\ref{sec:adequacy}, the adequacy of other model assumptions will be explored.
The simulated nozzle conditions will be compared to the ideal nozzle conditions, computed using Eq.~\ref{eq:soe_th}, to characterize the injector flows.
The first case discusses a choked flow injection of \SI{40}{\bar} argon into \SI{4}{\bar} argon. The needle lift is set at \SI{0.1}{\milli\m} such that the geometrical throat of the flow is located in the nozzle. This case is labeled `Base'.
Argon is chosen because of its relatively low sonic velocity (\SI{326}{\m\per\s} at \SI{40}{\bar} and \SI{300}{\K}) compared to methane and hydrogen (\SI{440}{\m\per\s} and \SI{1352}{\m\per\s}, respectively, at the same conditions). This significantly reduces the computational costs of the CFD simulations, since time integration is performed using a variable time-step based on the convective CFL number. 

It is not ensured that the throat of the flow is always located in the nozzle. Often in experiments concerning high-pressure gas injections, a (modified) liquid fuel injector such as a gasoline direct-injection (GDI) injector is employed \citep{peters2024characterizing,yip2020visualization, baert2010direct,welch2008challenges}. It is not uncommon that the nominal mass flow rates of these injectors are controlled by its needle lift. Therefore, often the throat is at the needle seat, depending on the actual lift height, and not at the nozzle exit. 
Note that the aim of the study is to relate reservoir conditions to the nozzle exit area. To test the adequacy of the 0D model in case the geometrical throat is not in the nozzle, the needle lift is decreased to \SI{0.040}{\milli\m} to locate the throat at the needle seat. All other settings are the same as the base setup. This case is labeled `Throttled'.

The two previously cases concern choked flow, as their upstream-to-downstream pressure ratio of 10 clearly exceed the critical ratio of $\sim$2. A choked flow injection has the advantage that the nozzle conditions are independent of downstream conditions, which allows to predict the mass flow rate. 
However, it is certainly not required to have choked flow in an engine application. 
Therefore, it is relevant to check the validity of the 0D model for a subsonic injector flow as well. This is done by increasing the chamber pressure up to \SI{30}{\bar}. This leads to an upstream-to-downstream pressure ratio of 1.33, and will therefore result in a subsonic injection. This case is labeled `Subsonic'.
Relevant injection conditions are provided in Table~\ref{tab:sim_conds}.
\begin{table}
\centering
\caption{Injection conditions of the simulated cases.}
\label{tab:sim_conds}
\begin{tabular}{@{}lccc@{}}
\toprule
Case & Base &  Throttled & Subsonic \\ \midrule
$P_\mathrm{res}$ $[$\SI{}{\bar}$]$ & 40 & 40 & 40 \\
$P_\mathrm{ch}$ $[$\SI{}{\bar}$]$ & 4 & 4 & 30 \\
$T_\mathrm{res}$ $[$\SI{}{\K}$]$ & 300 & 300 & 300 \\
$T_\mathrm{ch}$ $[$\SI{}{\K}$]$ & 300 & 300 & 300 \\
$T_\mathrm{wall}$ $[$\SI{}{\K}$]$ & 300 & 300 & 300 \\ 
Needle lift $[$\SI{}{\milli\m}$]$ & 0.10 & 0.04 & 0.10 \\ \bottomrule
\end{tabular}
\end{table}

Injector characteristics $C_\mathrm{D}$ and $C_\mathrm{M}$ are evaluated case-by-case by considering area-averaged nozzle exit plane values in the steady-state injection, which approximately starts after the needle is fully retracted. As a consequence, $C_\mathrm{D} \dot{m}_\mathrm{id}$ and $C_\mathrm{M} \dot{M}_\mathrm{id}$ in the 0D model are equal to the simulated nozzle mass and momentum flow rates. Therefore, the density and velocity are equal to the simulated values as well, see Eqs.~\ref{eq:U_expl} and \ref{eq:rho_expl}. For this reason, the assessment of the 0D model focuses on the nozzle pressure and temperature.

\subsection{Base \label{sec:sh_inj}}
\begin{figure}
    \centering
    \begin{subfigure}{0.47\textwidth}
         \centering
         \includegraphics[width=\textwidth]{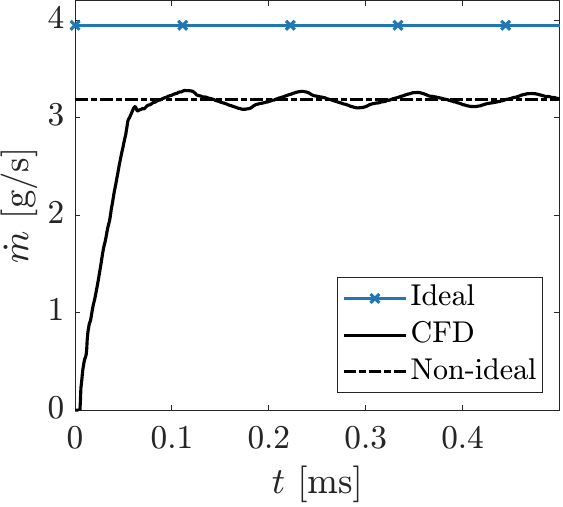}
     \end{subfigure}
     \hfill
     \begin{subfigure}{0.47\textwidth}
         \centering
         \includegraphics[width=\textwidth]{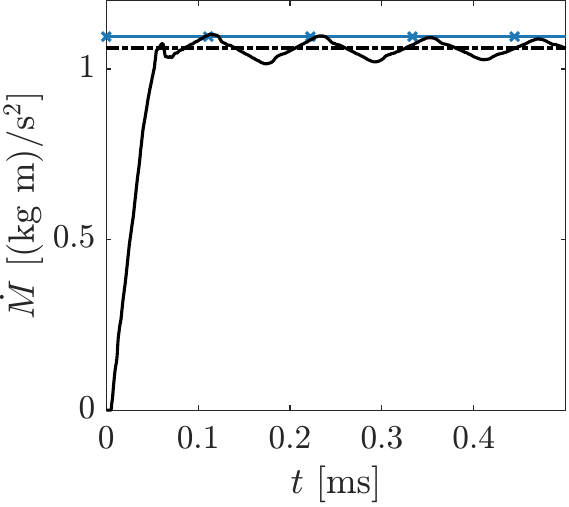}
     \end{subfigure}
    \caption{Area-averaged nozzle exit mass and momentum flow rates of the base CFD simulation, including its mean during the steady injection period and ideal values.}
    \label{fig:base_mdot_Mdot}
\end{figure}

\begin{figure}
    \centering
    \begin{subfigure}{0.47\textwidth}
         \centering
         \includegraphics[width=\textwidth]{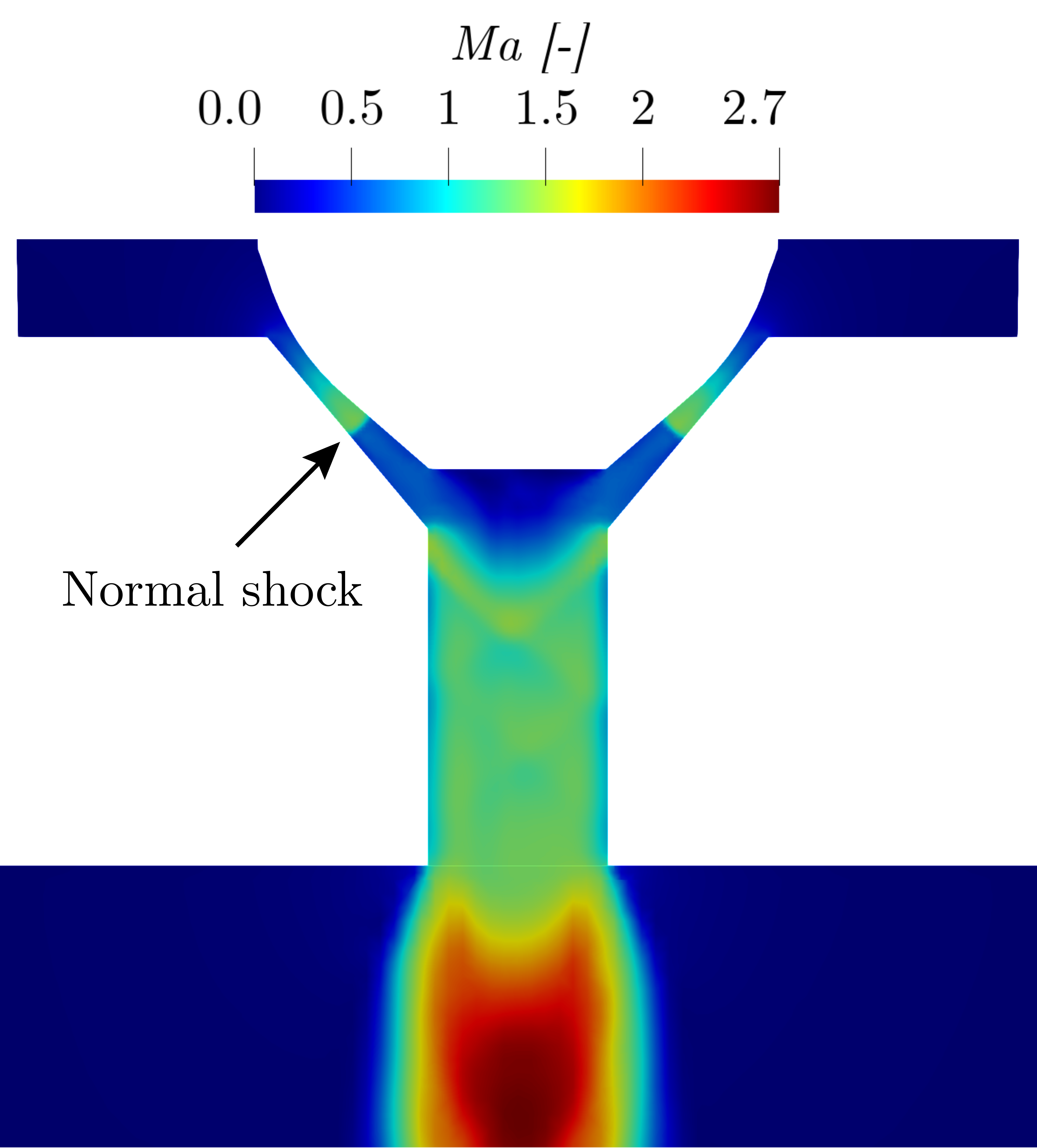}
         \caption{$t=\SI{0.05}{\milli\s}$}
     \end{subfigure}
     \hfill
     \begin{subfigure}{0.47\textwidth}
         \centering
         \includegraphics[width=\textwidth]{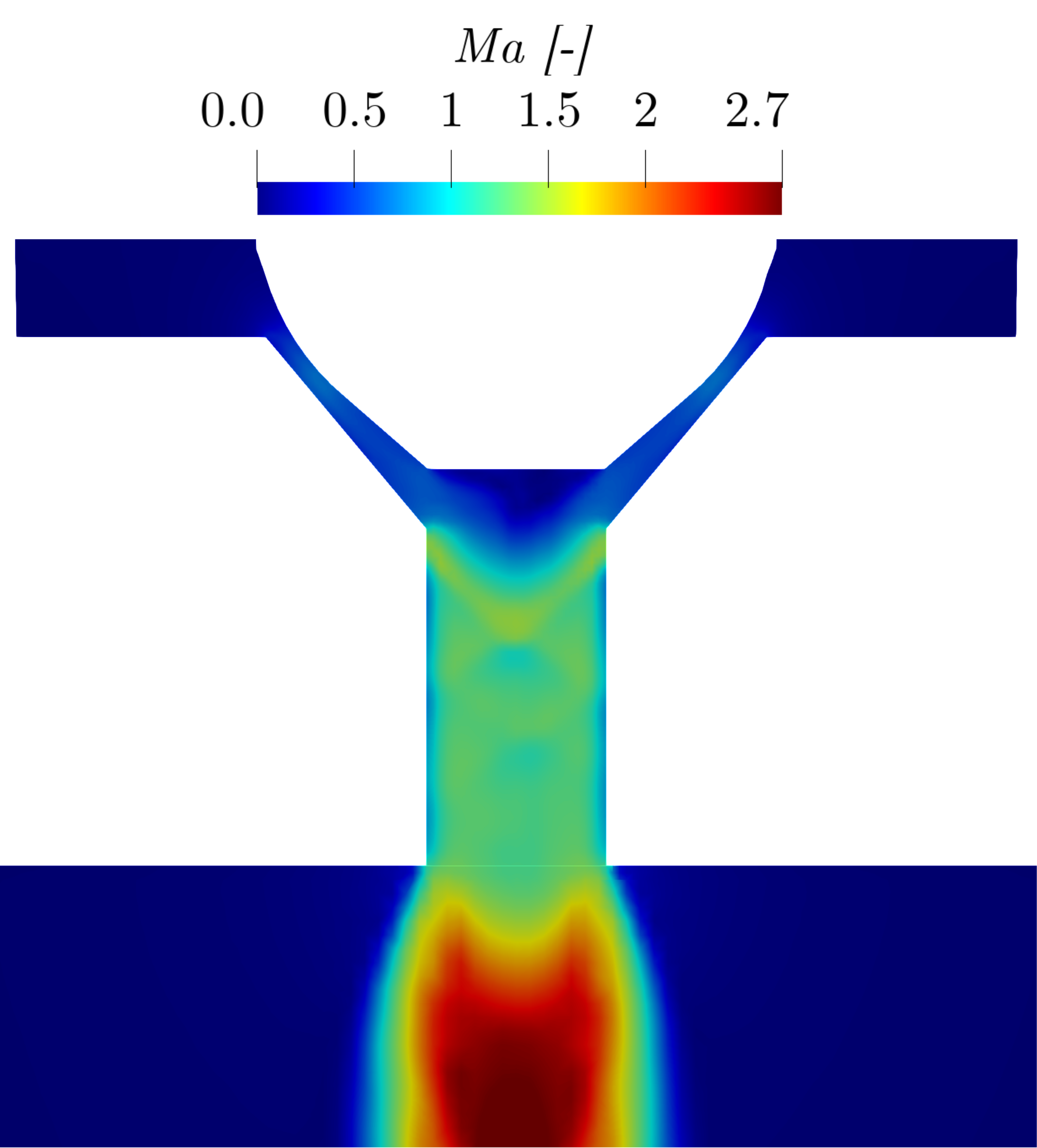}
          \caption{$t=\SI{0.06}{\milli\s}$}
     \end{subfigure}
    \caption{2D contour showing Mach number before and after the throat shifts to the nozzle.}
    \label{fig:base_Mach_contours}
\end{figure}

\noindent Figure~\ref{fig:base_mdot_Mdot} shows the area-averaged nozzle exit mass and momentum flow rates of the base case as function of time in black.
Both properties grow linearly during the first period, indicating that the injection rates are controlled by the needle lift. At $\sim$\SI{0.06}{\milli\s}, the linear increase transitions to the quasi steady period, during which the mean flow values fluctuate around a steady value. Note that the needle rise time is \SI{0.1}{\milli\s}, so it is not due to the stop of the needle motion. The end of the linear phase at \SI{0.06}{\milli\s} indicates that at that point in time the throat shifts from the needle seat to the nozzle exit, which is evident from the Mach contour plots just before and after the transition in Fig.~\ref{fig:base_Mach_contours}. Notably, at $t=\SI{0.05}{\milli\s}$, the converging-diverging geometry in the needle seat results in a supersonic flow beyond the throat, which eventually leads to the formation of a normal shock. 
The fluctuation at the nozzle exit is a result of a standing wave inside the injector body. 
Note that the period of the wave is about \SI{0.12}{\milli\s}, corresponding to a wavelength that is equal to the length of the injector body.

\begin{figure}
    \centering
    \begin{subfigure}{0.47\textwidth}
         \centering
         \includegraphics[width=\textwidth]{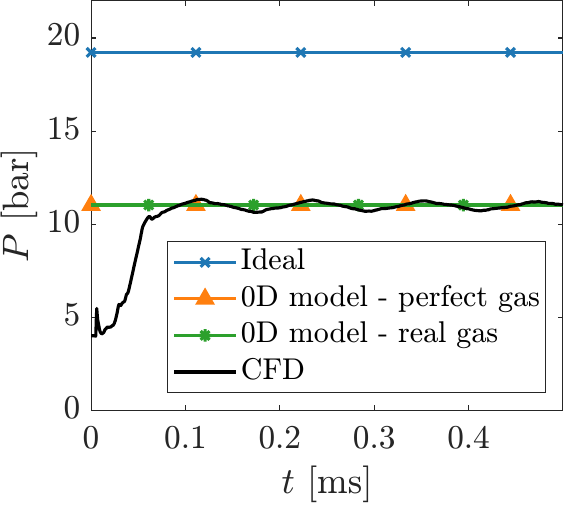}
     \end{subfigure}
     \hfill
     \begin{subfigure}{0.47\textwidth}
         \centering
         \includegraphics[width=\textwidth]{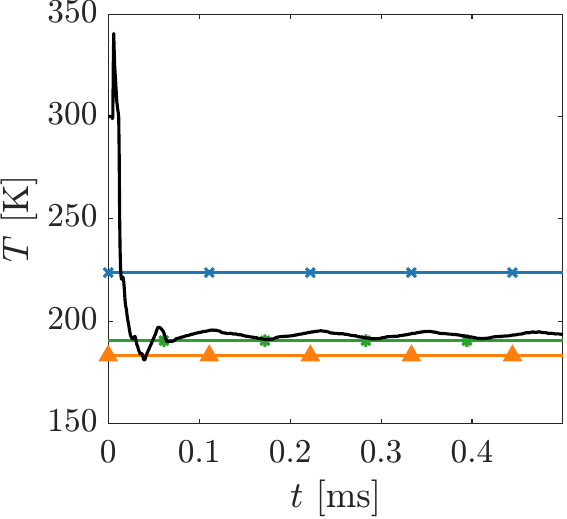}
     \end{subfigure}
    \caption{Simulated and calculated nozzle exit pressure and temperature of the base CFD simulation.}
    \label{fig:base_PTU}
\end{figure}

The blue lines in Fig.~\ref{fig:base_mdot_Mdot} denote the ideal nozzle flow rates, which are computed with Eq.~\ref{eq:soe_th}. The non-ideal nozzle flow rates are obtained by the mean values between 0.1 and \SI{0.5}{\milli\s} and are shown in orange. The resulting $C_\mathrm{D}$ and $C_\mathrm{M}$ values are provided in Table~\ref{tab:injchar_ar}. 
\begin{table}
\centering
\caption{$C_\mathrm{D}$ and $C_\mathrm{M}$ values for the argon injection simulations employing a single-hole injector}
\label{tab:injchar_ar}
\begin{tabular}{@{}lll@{}}
\toprule
Case      & $C_\mathrm{D}$ [-]   & $C_\mathrm{M}$ [-]   \\ \midrule
Base      & 0.81 & 0.96 \\
Throttled & 0.37 & 0.47 \\
Subsonic  & 0.67 & 0.36 \\ \bottomrule
\end{tabular}
\end{table}
It is noteworthy to mention that the ratio of $C_\mathrm{M}$ and $C_\mathrm{D}$ is the Mach number, as can be seen by substitution of Eqs.~\ref{eq:cd}, \ref{eq:cm}, \ref{eq:mdot0D}, and \ref{eq:Mdot0D}:
\begin{equation}\label{eq:cdcmmach}
    C_\mathrm{M} = \frac{\dot{M}}{\dot{M}_\mathrm{id}} = \frac{\left(\rho A U^2\right)}{\left(\rho A U^2\right)_\mathrm{id}} = \frac{\dot{m}}{\dot{m}_\mathrm{id}}\frac{U}{U_\mathrm{id}} = C_\mathrm{D}\frac{U}{c} = C_\mathrm{D} \mathit{Ma},
\end{equation}
where $c$ is the speed of sound. It shows that $\mathit{Ma}>1$ if $C_\mathrm{M}>C_\mathrm{D}$.
Fig.~\ref{fig:base_PTU} shows the nozzle exit pressure and temperature. The values of the steady pressure and temperature are lower than the ideal values, which is because the flow expanded beyond the sonic condition.
The largest deviation between the CFD result and the ideal value is observed for the pressure, which differ a factor of two, approximately. This shows that assuming an ideal injector is inappropriate to simulate a realistic injection. 

On the other hand, the 0D model with real gas thermodynamics (Eq.~\ref{eq:soe_isothermal}, plotted in green) shows excellent agreement with the simulated conditions. It is capable to determine the pressure and temperature at the nozzle exit with an accuracy of \SI{0.3}{\percent} and \SI{1.5}{\percent} for pressure and temperature, respectively.
Assuming a perfect gas (orange) results in a reasonable agreement with the simulated conditions. The largest difference is observed for the temperature, which results in an under-prediction of $\sim$\SI{5.3}{\percent}. The ratio of specific heats, which is needed for Eq.~\ref{eq:expl_T_pg_isenth}, is determined using CoolProp and at reservoir condition.
More precise values are provided in Table~\ref{tab:sh_devs}.

\subsection{Throttled case}
\begin{figure}
    \centering
    \includegraphics[width=0.7\linewidth]{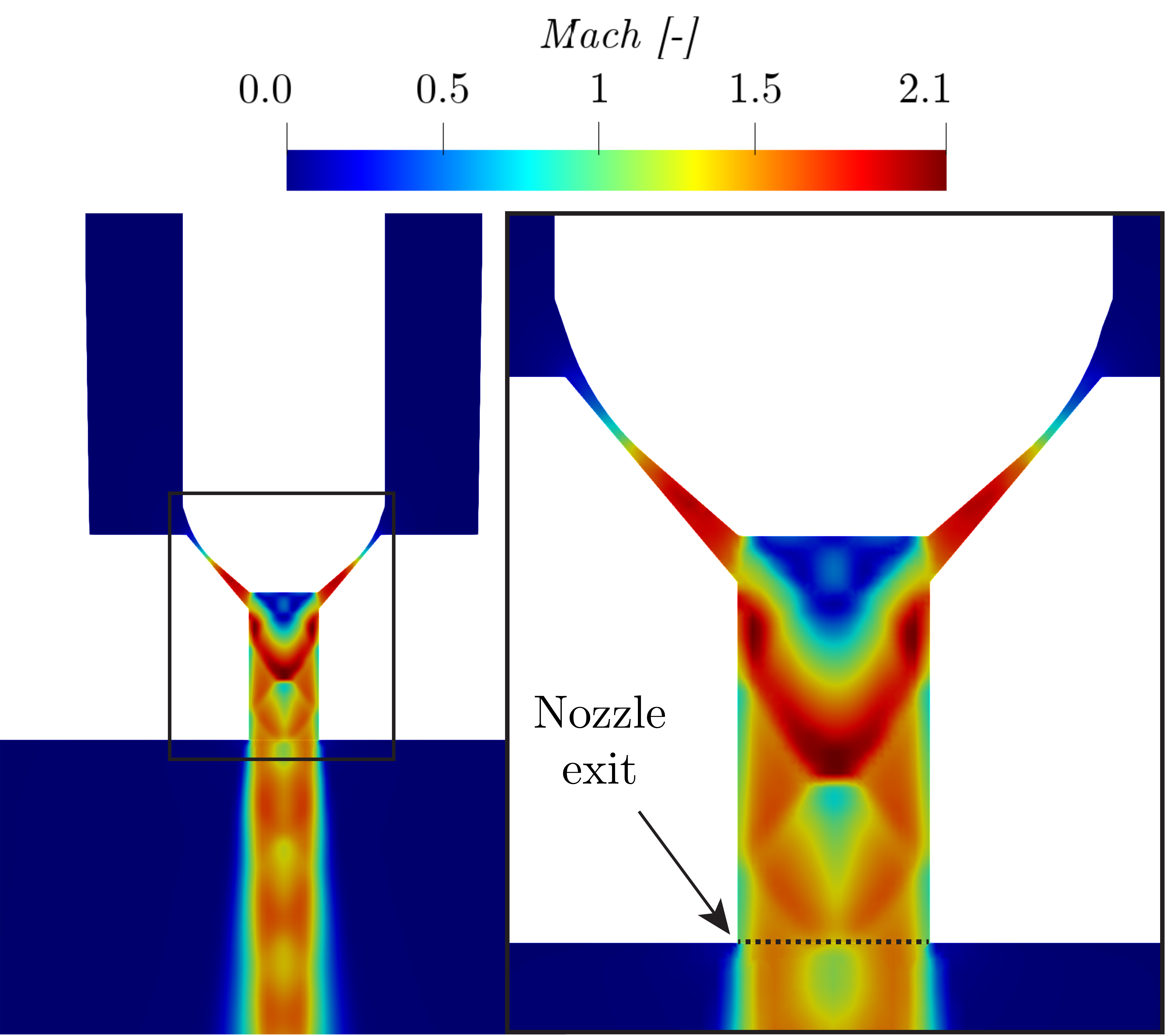}
    \caption{2D contour showing velocity magnitude of the throttled case at \SI{0.5}{\milli\s}.}
    \label{fig:throttled_contour}
\end{figure}

\noindent Figure~\ref{fig:throttled_contour} shows a 2D contour plot of local Mach number for the case with a smaller needle lift. The flow reaches Mach one at the needle seat, which confirms that the throat is now located there instead of in the nozzle exit. Due to the converging-diverging geometry around the needle seat, the fluid accelerates to beyond Mach one after it has passed the needle seat.

Due to the smaller throat of the flow, the steady mass flow rate is significantly lower compared to the base case. 
Since $C_\mathrm{D}$ and $C_\mathrm{M}$ are defined by the ideal conditions at the nozzle exit, they are also significantly lower than the base case (see Table~\ref{tab:injchar_ar}).
Still using these $C_\mathrm{D}$ and $C_\mathrm{M}$ values, the nozzle exit conditions are predicted accurately when accounting for real gas thermodynamics. The maximum deviation with respect to the CFD results is \SI{1.3}{\percent}, see Table~\ref{tab:sh_devs}. Assuming a perfect gas leads to larger differences.

\subsection{Subsonic}
\noindent The last single-hole injector setup simulates a subsonic injection. As expected from theory~\citep{shapiro1953dynamics}, the nozzle exit pressure (\SI{30}{\bar}) is approximately equal to the downstream pressure, see Table~\ref{tab:sh_devs}.
It can also be concluded from that table that the 0D model prediction is in excellent agreement with the CFD simulation, with a maximum difference of \SI{0.5}{\percent}. 
The perfect gas model performs well too, with a maximum difference of \SI{2.2}{\percent}.
Comparing the performances of the 0D model in the choked injections with the subsonic case, it is observed that it performs better in the latter. This can be explained by the fact that the decrease in fluid temperature of the subsonic injection is less due to the smaller expansion, leading to less heat transfer to the injector walls.
\begin{table}
\centering
\caption{Nozzle exit pressures and temperatures of the single-hole injector flows. The percentages between brackets indicate the difference between the isothermal GCM and the CFD result.}
\label{tab:sh_devs}
\begin{tabular}{@{}llll@{}}
\toprule 
Case                        & Model          & $P$ [\SI{}{\bar}] & $T$ [\SI{}{\K}] \\ \midrule
\multirow{3}{*}{Base}       & CFD              & 10.99            & 193.4 \\
                            & 0D - perfect gas & 11.01 ($+0.2$\%) & 183.2 ($-5.3$\%) \\
                            & 0D - real gas    & 11.02 ($+0.3$\%) & 190.5 ($-1.5$\%) \\ \midrule
\multirow{3}{*}{Throttled}  & CFD              & 4.48             & 180.1 \\
                            & 0D - perfect gas & 4.30 ($-3.7$\%)  & 167.6 ($-6.9$\%) \\
                            & 0D - real gas    & 4.48 ($+0.2$\%)  & 177.7 ($-1.3$\%) \\ \midrule
\multirow{3}{*}{Subsonic}   & CFD              & 30.15            & 277.8 \\
                            & 0D - perfect gas & 30.88 ($+2.2$\%) & 277.2 ($-0.2$\%) \\
                            & 0D - real gas    & 30.30 ($+0.5$\%) & 278.8 ($+0.4$\%) \\ \midrule
\end{tabular}
\end{table}

\section{Case studies at engine conditions\label{sec:case_studies}}
\noindent The previous section considered flow bench conditions, in which all wall and initial fluid temperatures were assumed to be equal. At these conditions, the isothermal 0D model was shown to be accurate. However, in an engine, the temperature of the fuel entering the injector ($\pm$\SI{313}{\K}) will be considerably lower than the injector wall temperature ($\pm$\SI{353}{\K}). It is expected that heat is transferred from the injector walls to the fuel. Thus, the adequacy of the isothermal assumption made in Eq.~\ref{eq:soe_isothermal} is not ensured. 
In this section, case studies are performed for two different injection strategies: an LPDI injection employing the poppet-valve type injector, and a HPDI injection employing the multi-hole injector. In both strategies, injections of methane and hydrogen are simulated.

The effect of heat transfer is assessed by comparing simulations at flow bench and engine conditions. In the flow bench simulations, the initial and boundary conditions for temperature are set to \SI{300}{\K}. 
For the simulations at engine conditions, the initial and boundary conditions for the temperature are chosen to mimic engine conditions. Note that these temperatures can vary depending on the specific layout of the fuel injection system and engine design.
The injector wall temperature is estimated slightly above the coolant temperature and set to \SI{353}{\K}. The temperature of the fuel entering the injector is set to \SI{313}{\K}.
Furthermore, it is assumed that during engine operation the fuel heats up in between injections until it reaches the injector wall temperature. Hence, the initial temperature of the fuel inside the injector is also set to \SI{353}{\K}.
For the LPDI cases, the injection pressure is \SI{40}{\bar}. The chamber pressure is set to \SI{4}{\bar}. The temperature in the chamber is set to \SI{405}{\K}, which is calculated from an isentropic compression of air starting at \SI{2}{\bar} and \SI{333}{\K}, which can be considered as relevant intake conditions. 
In the HPDI setups injections of methane and hydrogen are performed at \SI{300}{\bar}. The chamber conditions are \SI{60}{\bar} and \SI{1200}{\K}. 
Note that the chamber conditions are chosen to mimic engine conditions, but will not affect the internal injector flow since it concerns choked flow injections.
The injection conditions for the cases with flow bench temperatures, referred to as FB, and with engine temperatures, referred to as ICE, are summarized in Table~\ref{tab:sim_conds_case_studies}.
\begin{table}
\centering
\caption{Injection conditions of the LPDI and HPDI case studies. Abbreviations FB and ICE indicate that the cases are performed at flow bench and engine temperatures, respectively.}
\label{tab:sim_conds_case_studies}
\small
\begin{tabular}{@{}lcccccccc@{}}
\toprule
 & \multicolumn{4}{c}{\textbf{LPDI (poppet-valve)}} &
  \multicolumn{4}{c}{\textbf{HPDI (multi-hole)}} \\ \cmidrule(rl){2-5} \cmidrule(rl){6-9}
 & FB & ICE & FB & ICE & FB & ICE & FB & ICE \\ \midrule
Species $[$-$]$                     & \multicolumn{2}{c}{CH$_4$} & \multicolumn{2}{c}{H$_2$} & \multicolumn{2}{c}{CH$_4$} & \multicolumn{2}{c}{H$_2$} \\
$P_\mathrm{res}$ $[$\SI{}{\bar}$]$  & \multicolumn{2}{c}{40} & \multicolumn{2}{c}{40} &  \multicolumn{2}{c}{300} & \multicolumn{2}{c}{300} \\
$P_\mathrm{ch}$ $[$\SI{}{\bar}$]$   & \multicolumn{2}{c}{4} & \multicolumn{2}{c}{4} &   \multicolumn{2}{c}{60} & \multicolumn{2}{c}{60} \\
$T_\mathrm{res}$ $[$\SI{}{\K}$]$    & 300 & 313 & 300 & 313 & 300 & 313 & 300 & 313 \\
$T_\mathrm{ch}$ $[$\SI{}{\K}$]$     & 300 & 405 & 300 & 405 & 300 & 1200 & 300 & 1200 \\
$T_\mathrm{wall}$ $[$\SI{}{\K}$]$   & 300 & 353 & 300 & 353 & 300 & 353 & 300 & 353 \\ \bottomrule
\end{tabular}
\end{table}

\subsection{Low-pressure direct-injection \label{sec:lpdi}}
\begin{figure}
    \centering
    \includegraphics[width=0.6\linewidth]{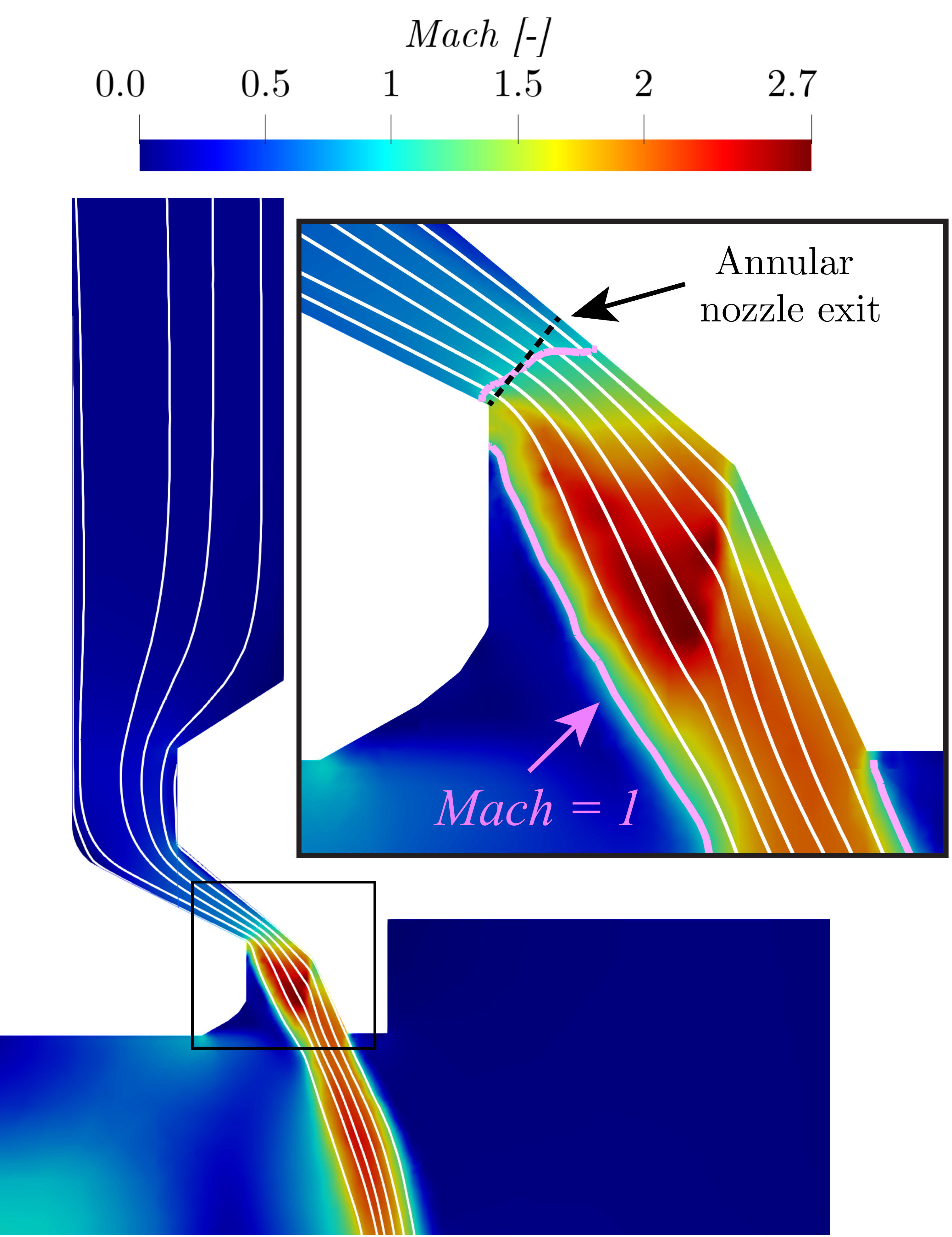}
    \caption{2D contour plot showing Mach number at \SI{0.4}{\milli\s} of the hydrogen injection through the poppet-valve type injector. The white curves depict streamlines.}
    \label{fig:lpdi_contour}
\end{figure}

\noindent The LPDI injections employ the poppet-valve type injector shown in Fig.~\ref{fig:oo_domain}. Starting with the discharge and momentum coefficients from Table~\ref{tab:injchar_lpdi}, it is observed that the injector behaves close to ideal, because the $C_\mathrm{D}$ and $C_\mathrm{M}$ values approach one. 
Furthermore, it can be concluded that mean nozzle exit velocity is subsonic ($C_\mathrm{M}<C_\mathrm{D}$, see Eq.~\ref{eq:cdcmmach}) despite the fact that the cases satisfy the choked flow criterion. This can be explained by the Mach field, which is shown for the FB hydrogen case in Fig.~\ref{fig:lpdi_contour}.\@ It is clear that the velocity across the nozzle exit is non-uniform. Due to the specific injector geometry, a part of the flow reaches Mach one after it has passed the nozzle exit plane. 
\begin{table}
\centering
\caption{$C_\mathrm{D}$ and $C_\mathrm{M}$ values obtained from the CFD simulations for the LPDI cases employing a poppet-valve type injector}
\label{tab:injchar_lpdi}
\begin{tabular}{@{}cccc@{}}
\toprule
\multicolumn{2}{c}{Case}  & $C_\mathrm{D}$ [$-$] & $C_\mathrm{M}$ [$-$]   \\ \midrule
\multirow{2}{*}{Methane}  & FB & 0.98 & 0.93 \\
                          & ICE  & 0.97 & 0.93 \\ \midrule
\multirow{2}{*}{Hydrogen} & FB & 0.97 & 0.93 \\
                          & ICE  & 0.97 & 0.93 \\ \midrule 
\end{tabular}
\end{table}

\begin{figure}
    \centering
    \begin{subfigure}{0.32\textwidth}
         \centering
         \includegraphics[width=\textwidth]{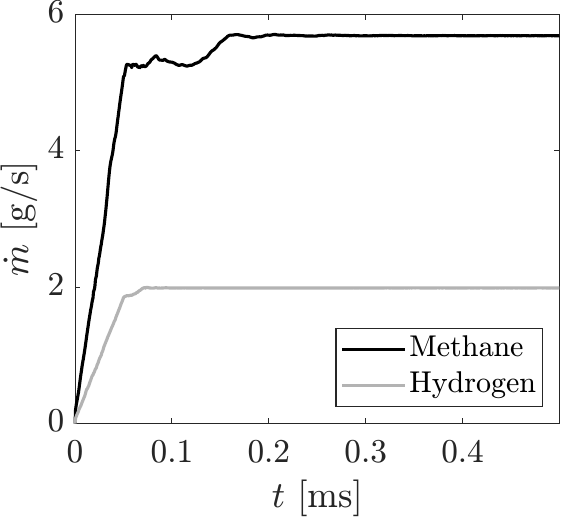}
     \end{subfigure}
     \hfill
     \begin{subfigure}{0.32\textwidth}
         \centering
         \includegraphics[width=\textwidth]{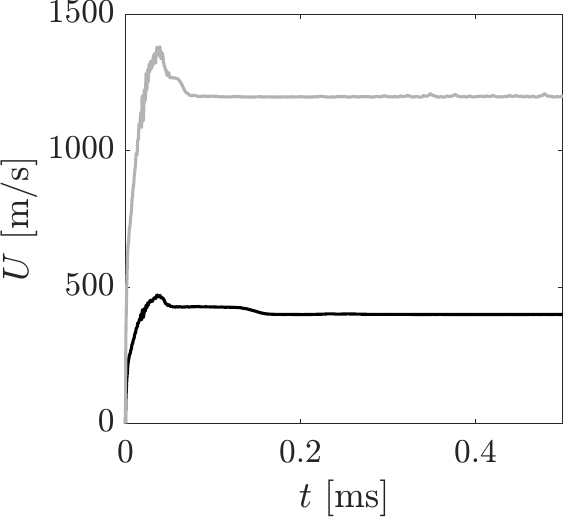}
     \end{subfigure}
     \hfill
     \begin{subfigure}{0.32\textwidth}
         \centering
         \includegraphics[width=\textwidth]{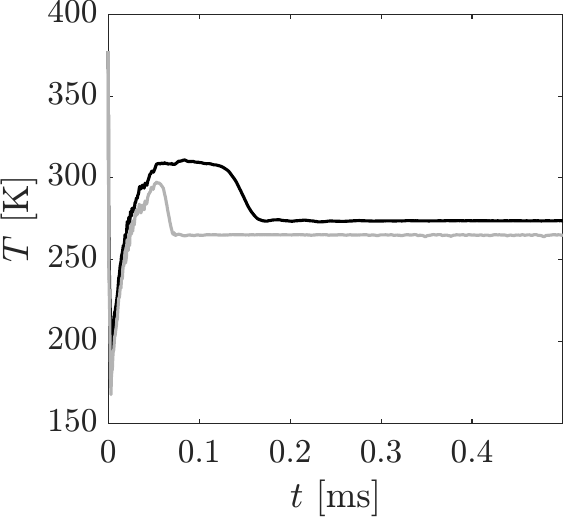}
     \end{subfigure}
    \caption{Nozzle exit mass flow rate, velocity and temperature vs.\@ time for the methane and hydrogen LPDI injections at engine temperatures with $P_\mathrm{res}=$ \SI{40}{\bar} and $T_\mathrm{res}=$ \SI{313}{\K}.}
    \label{fig:lpdi_ice}
\end{figure}

Figure~\ref{fig:lpdi_ice} shows the simulated nozzle exit conditions for the methane and hydrogen injections at engine conditions. Around \SI{0.15}{\milli\s} and \SI{0.065}{\milli\s} in the methane and hydrogen profiles, respectively, they change value. This is especially evident in the methane temperature profile and is due to hot fluid (\SI{353}{\K}) initially present in the injector that leaves the injector. Afterward, the injector only contains fluid originating from the upstream inflow boundary (\SI{313}{\K}).

The nozzle exit conditions determined by the isothermal 0D model with real gas thermodynamics are shown in Table~\ref{tab:lpdi_results}. It is good to note that the discharge and momentum coefficients for the ICE cases are determined after the fuel initially present in the injector has left the nozzle, i.e.\@ after \SI{0.2}{\milli\s}. 
The 0D model results are in good agreement with the simulated values and deviate no more than \SI{1.5}{\percent}, despite the temperature of the injector walls being higher than the fluid entering the domain. Although this temperature difference is \SI{40}{\K}, the nozzle exit temperature only increases with $\sim$\SI{14}{\K}. The differences in pressure are negligible. 
Assuming perfect gas behavior leads to larger differences, up to \SI{6.5}{\percent} in pressure. It demonstrates that real gas behavior is preferred for this setup, despite the relatively low injection pressure of \SI{40}{\bar}.
\begin{table}
\centering
\caption{Nozzle exit pressure and temperature of the LPDI cases employing the poppet-valve type injector. The percentages between brackets indicate the difference between the isothermal GCM and the CFD result.}
\label{tab:lpdi_results}
\begin{tabular}{@{}cllll@{}}
\toprule
\multicolumn{2}{c}{Case}    & Model             & $P$ [\SI{}{\bar}] & $T$ [\SI{}{\K}] \\ \midrule
\multirow{6}{*}{Methane}    & \multirow{3}{*}{FB} & CFD               & 22.26             & 260.1 \\
                            &                       & 0D, perfect gas   & 23.72 ($+$6.5\%)  & 257.0 ($-$1.2\%) \\
                            &                       & 0D, real gas      & 22.60 ($+$1.5\%)  & 261.2 ($+$0.4\%) \\ \cmidrule(l){2-5} 
                            & \multirow{3}{*}{ICE}  & CFD               & 22.21             & 273.8 \\
                            &                       & 0D, perfect gas   & 23.26 ($+$4.7\%)  & 268.8 ($-$1.8\%) \\
                            &                       & 0D, real gas      & 22.35 ($+$0.6\%)  & 272.7 ($-$0.4\%) \\ \midrule
\multirow{6}{*}{Hydrogen}   & \multirow{3}{*}{FB} & CFD               & 21.33             & 251.9 \\
                            &                       & 0D, perfect gas   & 21.24 ($-$0.4\%)  & 251.6 ($-$0.1\%) \\
                            &                       & 0D, real gas      & 21.62 ($+$1.4\%)  & 252.6 ($+$0.3\%) \\ \cmidrule(l){2-5} 
                            & \multirow{3}{*}{ICE}  & CFD               & 21.36             & 265.2 \\
                            &                       & 0D, perfect gas   & 21.10 ($-$1.2\%)  & 262.5 ($-$1.0\%) \\
                            &                       & 0D, real gas      & 21.46 ($+$0.5\%)  & 263.5 ($-$0.6\%) \\ \bottomrule
\end{tabular}
\end{table}

\subsection{High-pressure direct-injection \label{sec:hpdi}}
\noindent This section discusses the HPDI cases employing the multi-hole injector shown in Fig.~\ref{fig:mh_domain}. Similar to previous section, the injections are simulated with FB temperatures of \SI{300}{\K} and with engine temperatures, in which the injector walls and fluid in the injector have a higher temperature than the fluid entering the domain upstream. The injection conditions are provided in Table~\ref{tab:sim_conds_case_studies}. 

\begin{figure}
    \centering
    \begin{subfigure}{0.45\textwidth}
         \centering
         \includegraphics[width=\textwidth]{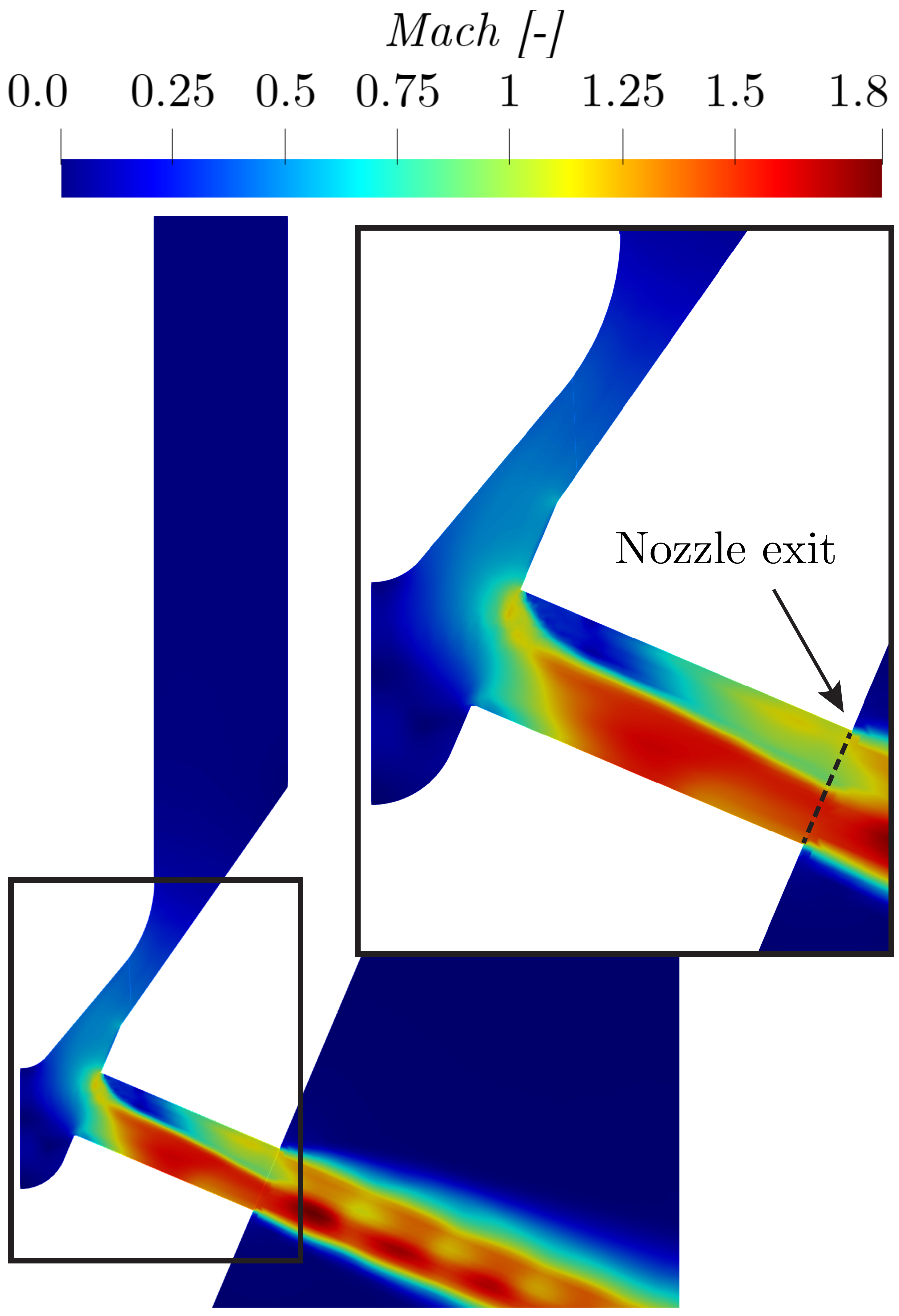}
         \caption{Center plane of injector.}
         \label{fig:hpdi_contours_injector}
     \end{subfigure}
     \hfill
     \begin{subfigure}{0.52\textwidth}
         \centering
         \includegraphics[width=\textwidth]{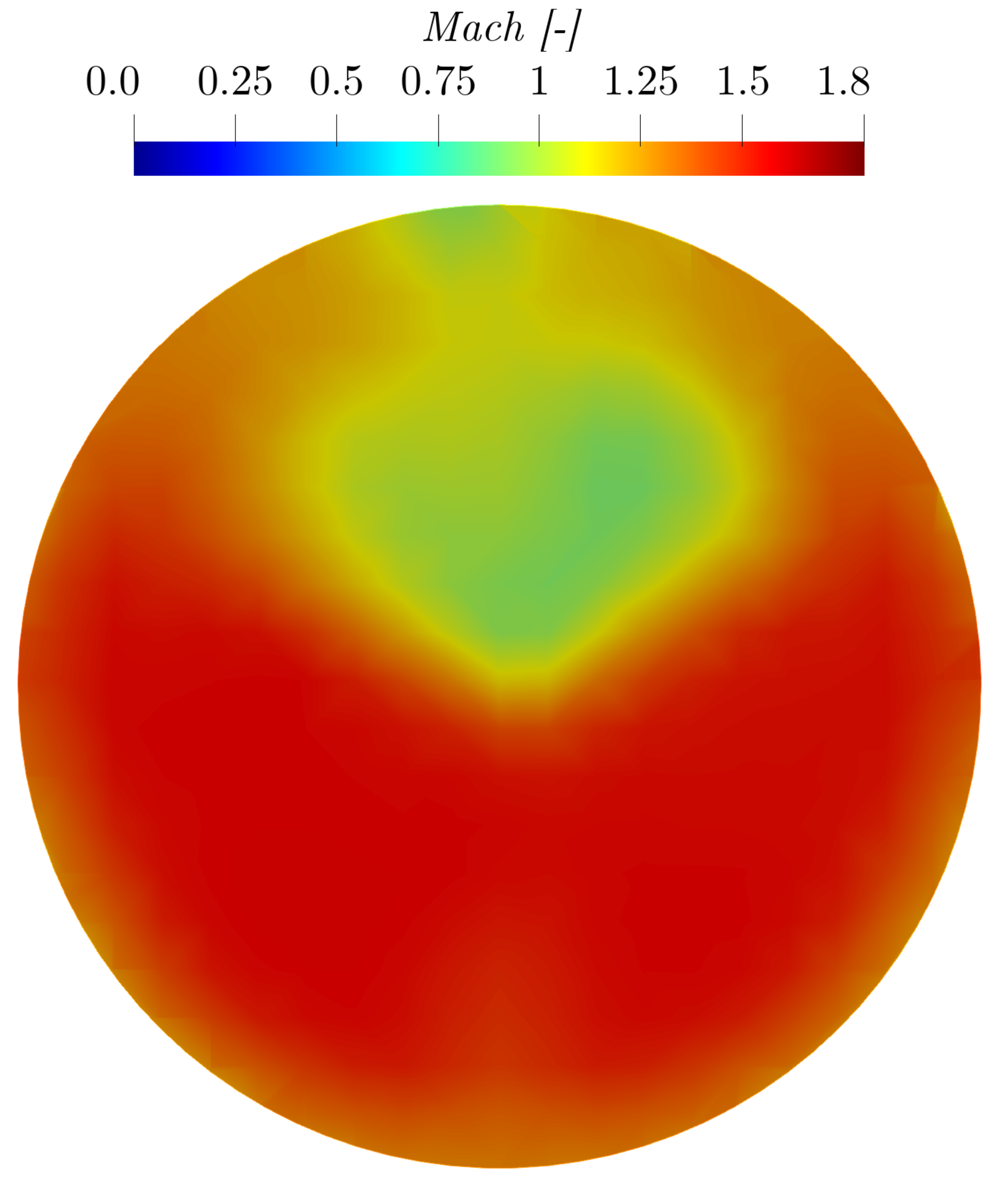}
         \caption{Nozzle exit plane}
         \label{fig:hpdi_contours_nozzle}
     \end{subfigure}
    \caption{2D contour plots of the hydrogen injection through the multi-hole injector showing velocity magnitude at $t=\SI{0.4}{\milli\s}$.}
    \label{fig:hpdi_contours}
\end{figure}

Figure~\ref{fig:hpdi_contours_injector} shows a 2D contour of local Mach number for the hydrogen injection with FB temperatures. Due to the sharp, right angle between the sac and the nozzle, the flow field in the nozzle is highly non-uniform. This non-ideal flow behavior results in a discharge coefficient of approximately 0.8 and a momentum coefficient of 0.8 for methane and 0.9 for hydrogen, see Table~\ref{tab:injchar_hpdi}. As for the LPDI cases, the values for $C_\mathrm{M}$ and $C_\mathrm{D}$ are determined based on the flow rates after the hot gas initially present in the injector has left the nozzle.
In Fig.~\ref{fig:hpdi_contours_nozzle}, the velocity magnitude field associated with the local Mach number field at the nozzle exit plane shows values ranging from \SI{1100}{\m\per\s} up to \SI{1900}{\m\per\s}. The mean value is supersonic, which is why $C_\mathrm{M}>C_\mathrm{D}$ for all cases.
Another consequence of the non-ideal flow behavior is that the simulated nozzle pressures and temperatures are significantly lower than the ideal values, which are shown in italic in Table~\ref{tab:hpdi_results}. Especially the difference in pressure is significant, with its value roughly a factor two lower than the ideal value.
\begin{table}
\centering
\caption{$C_\mathrm{D}$ and $C_\mathrm{M}$ values for the HPDI cases employing the multi-hole injector.}
\label{tab:injchar_hpdi}
\begin{tabular}{@{}cccc@{}}
\toprule
\multicolumn{2}{c}{Case}  & $C_\mathrm{D}$ [$-$] & $C_\mathrm{M}$ [$-$]   \\ \midrule
\multirow{2}{*}{Methane}  & FB & 0.68 & 0.78\\
                          & ICE  & 0.68 & 0.80 \\ \midrule
\multirow{2}{*}{Hydrogen} & FB & 0.71 & 0.88 \\
                          & ICE  & 0.71 & 0.87 \\ \midrule 
\end{tabular}
\end{table}

The discharge coefficient is insensitive to the increased injector wall temperature, and the momentum coefficient increased only slightly. This indicates that the effect of heat transfer on the mass and momentum flow rate is minor for the simulated high-pressure, supersonic injections. Moreover, it shows that the $C_\mathrm{D}$ and $C_\mathrm{M}$ values obtained at flow bench conditions can be used to determine nozzle exit conditions at engine-conditions as well even though wall temperatures might differ.
The differences between the FB and ICE cases in nozzle pressure and temperature are approximately \SI{4}{\bar}, \SI{15}{\K} for the methane injections and \SI{1}{\bar}, \SI{15}{\K} for the hydrogen injections. These differences are caused by the different temperature of the incoming fluid.

Despite the large differences in velocity across the nozzle exit plane and the pronounced non-ideal injector behavior, the isothermal 0D model with real gas thermodynamics is able to estimate the mean nozzle exit conditions of the FB cases within \SI{8}{\percent} and \SI{4}{\percent} for the ICE cases, see Table~\ref{tab:hpdi_results}. Clearly, a perfect gas assumption is not adequate for the high-pressure methane injections, with differences leading up to \SI{48}{\percent}. Also for the hydrogen injections, the perfect gas model performs worse with differences exceeding \SI{10}{\percent}.
This indicates that the real gas behavior of methane and hydrogen needs to be taken into account when calculating nozzle exit conditions for high pressure injections.

\begin{table}
\centering
\caption{Nozzle exit pressure and temperature of the HPDI cases employing the multi-hole injector. The percentages between brackets indicate the difference between the isothermal GCM and the CFD result.}
\label{tab:hpdi_results}
\small
\begin{tabular}{@{}lllll@{}}
\toprule
\multicolumn{2}{c}{Case}                           & Model             & $P$ [\SI{}{\bar}] & $T$ [\SI{}{\K}] \\ \midrule
\multirow{8}{*}{Methane}    & \multirow{4}{*}{FB}  & CFD               & 61.55 & 216.4 \\
                            &                       & \textit{Ideal}    & \textit{129.2} & \textit{251.5} \\
                            &                       & 0D - perfect gas   & 90.83 ($+$47.6\%) & 192.5 ($-$11.0\%) \\ 
                            &                       & 0D - real gas      & 66.51 ($+$8.1\%) & 222.4 ($+$2.8\%) \\ \cmidrule(l){2-5}
                            & \multirow{4}{*}{ICE}  & CFD               & 65.26 & 230.1 \\
                            &                       & \textit{Ideal}    & \textit{134.5} & \textit{262.6} \\
                            &                       & 0D - perfect gas   & 85.03 ($+$30.3\%) & 199.3 ($-$13.4\%) \\
                            &                       & 0D - real gas      & 67.61 ($+$3.6\%)  & 230.7 ($+$0.3\%)  \\ \midrule
\multirow{8}{*}{Hydrogen}   & \multirow{4}{*}{FB}  & CFD               & 71.36 & 215.5 \\
                            &                       & \textit{Ideal}    & \textit{149.0} & \textit{244.9} \\
                            &                       & 0D - perfect gas   & 64.86 ($-$9.1\%)  & 202.5 ($-$6.0\%) \\
                            &                       & 0D - real gas      & 71.26 ($-$0.1\%)  & 211.7 ($-$1.7\%) \\ \cmidrule(l){2-5} 
                            & \multirow{4}{*}{ICE}  & CFD               & 72.12 & 229.9 \\
                            &                       & \textit{Ideal}    & \textit{149.6} & \textit{256.0} \\
                            &                       & 0D - perfect gas   & 64.68 ($-$10.3\%) & 211.7 ($-$7.9\%) \\
                            &                       & 0D - real gas      & 70.85 ($-$1.8\%)  & 220.9 ($-$3.9\%) \\ \bottomrule
\end{tabular}
\end{table}


\section{Adequacy of different model assumptions\label{sec:adequacy}}
\noindent Thus far, the 0D model has been tested for three injector designs, choked and subsonic injections, different needle lift heights, high and low injection pressure, and with FB and ICE temperatures. The isothermal 0D model with real gas thermodynamics showed to perform well in all cases. Still, it is interesting to explore other assumptions, such as for an isenthalpic (\ref{eq:soe_isenthalpic}) and isentropic (\ref{eq:soe_isentropic}) assumption. The results are reported in Table~\ref{tab:assumps}. 

Overall, the isothermal and isenthalpic assumptions are approximately equally accurate, with an average error of $\sim$\SI{1.5}{\percent}. The isentropic assumption performs well for the poppet-valve type injector, which was shown to behave rather ideal, but fails to accurately predict nozzle conditions in the other cases. 
It can be concluded that, among the three tested assumptions, the isothermal and isenthalpic assumptions perform best.
\begin{table}
\centering
\small
\caption{Overview of differences in nozzle pressure and temperature between the simulations and 0D calculations for the hydrogen HPDI injection at engine conditions for various assumptions.}
\label{tab:assumps}
\setlength{\tabcolsep}{4pt}
\begin{tabular}{@{}cllllllll@{}}
\toprule
\multicolumn{3}{c}{\multirow{2}{*}{Case}} & \multicolumn{2}{c}{Isothermal} & \multicolumn{2}{c}{Isenthalpic} & \multicolumn{2}{c}{Isentropic} \\ \cmidrule(l){4-5} \cmidrule(l){6-7} \cmidrule(l){8-9} 
\multicolumn{3}{l}{} & $\Delta P$ [\%] & $\Delta T$ [\%] & $\Delta P$ [\%] & $\Delta T$ [\%] & $\Delta P$ [\%] & $\Delta T$ [\%] \\ \midrule
\multirow{3}{1.2cm}{Single-hole} & \multirow{3}{*}{\ce{Ar}} & Base & \(+0.3\) & \(-1.5\) & \(-0.8\) & \(-2.5\) & \(-10.9\) & \(-11.6\) \\
                                                  &  & Throttled & \(-0.2\) & \(-1.4\) & \(-5.1\) & \(-6.0\) & \(-47.8\) & \(-46.1\) \\
                                                  &  & Subsonic & \(+0.5\) & \(+0.4\) &\( +0.0\) & \(-0.1\) &\( -6.7 \)& \(-6.2\) \\ \midrule
\multirow{4}{1.2cm}{Poppet-valve} & \multirow{2}{*}{\ce{CH4}} & FB & \(+1.5\) & \(+0.4\) & \(+1.3\) & \(+0.3\) & \(+0.7\) &\( -0.3\) \\
                                                 &  & ICE & \(+0.6\) & \(-0.4\) & \(+0.5\) & \(-0.5\) & \(-0.4\) & \(-1.2\) \\ \cmidrule{2-9}
                     & \multirow{2}{*}{\ce{H2}}    & FB& \(+1.4\) & \(+0.3\) & \(+1.4 \)&\( +0.3\) &\( +0.4\) & \(-0.7\) \\
                                                 &  & ICE & +\(0.5\) & \(-0.6\) &\( +0.5\) & \(-0.6 \)& \(-0.7\) & \(-1.8\) \\ \midrule
\multirow{4}{1.2cm}{Multi-hole} & \multirow{2}{*}{\ce{CH4}} & FB & \(+8.1\) & \(+2.8\) & \(-1.2\) & \(-1.0\) & \(-16.1\) & \(-7.0\) \\
                                                 &  & ICE & \(+3.6\) &\( +0.3\) &\( -4.2\) & \(-3.4\) & \(-17.0\) & \(-9.3\) \\ \cmidrule{2-9}
                       & \multirow{2}{*}{\ce{H2}} & FB & \(-0.1\) & \(-1.7\) & \(+1.3\) & \(-0.3\) & \(-11.3 \)& \(-12.3\) \\
                                                 &  & ICE & \(-1.8 \)&\( -3.9 \)& \(-0.4 \)& \(-2.6 \)&\( -12.5\) & \(-14.1\) \\ \bottomrule 
\end{tabular}
\end{table}

\begin{figure}
    \centering
    \includegraphics[width=0.55\linewidth]{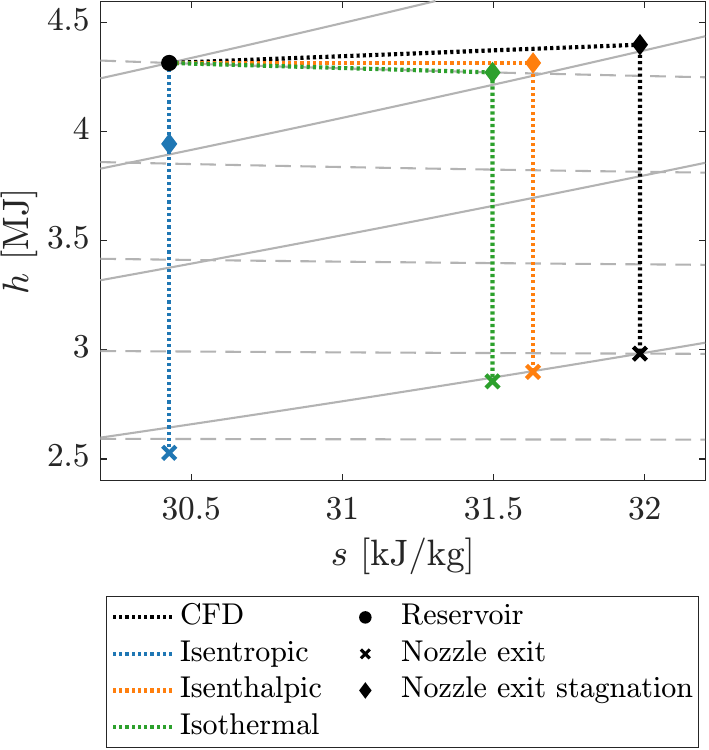}
    \caption{Enthalpy vs.\@ entropy diagram showing the conditions simulated by CFD and calculated with the real gas GCM. The solid lines denote isobars and the dashed lines are isotherms. They are distributed evenly in pressure and temperature, respectively.}
    \label{fig:hs_hpdi}
\end{figure}

To gain insight in the reasons of the model performances, the enthalpy-entropy diagram of the hydrogen HPDI ICE case is shown in Fig.~\ref{fig:hs_hpdi}. 
Note that the vertical drop between the nozzle exit and its stagnation state is $U^2/2$ in all cases, since velocity $U$ is directly derived from the mass and momentum flow rates.
Noticeably, both enthalpy and entropy increase between the reservoir and the stagnation state of the nozzle exit in the simulated data. 
The increase in entropy is expected due to irreversible processes and heat transfer to the fluid due the higher temperature of the injector walls which will inevitably exist due to the expansion of the gas in the injector. By the first law of thermodynamics, also the increase in enthalpy between the reservoir and stagnation condition at the nozzle exit state can be related to the heat transfer.

The isenthalpic model performs slightly better than the isothermal model in this specific case, which can be attributed to the increase in enthalpy between the reservoir and the stagnation nozzle exit condition and negative slope of the isotherm.

\section{Sensitivity of injector characteristics on nozzle conditions\label{sec:sensitivity}}
\noindent In previous chapters, the discharge and momentum coefficients were evaluated from detailed CFD simulations for each specific case. However, the goal of the 0D model is to be able to simulate any reservoir injection condition using experimentally determined discharge and momentum coefficients. These coefficients are typically only measured at a few conditions, such that inter- or extrapolation is needed to get the specific values at the desired condition. Besides, experimental values always come with a confidence interval. For these two reasons, the sensitivity of the discharge and momentum coefficients on nozzle conditions is studied. 

\begin{figure}
    \centering
        \begin{subfigure}{0.32\textwidth}
         \centering
         \includegraphics[width=\textwidth]{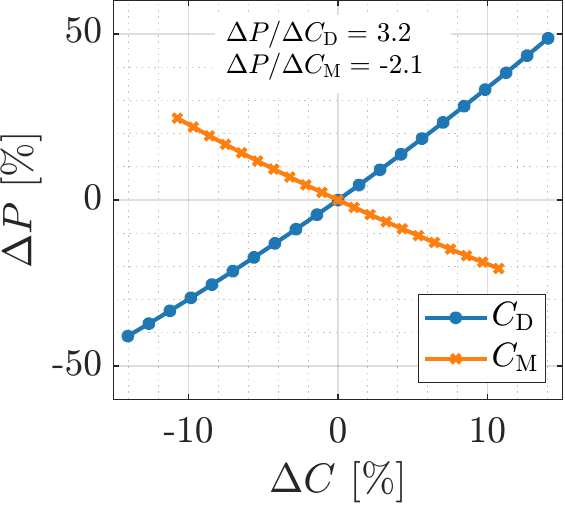}
     \end{subfigure}
     \hfill
     \begin{subfigure}{0.32\textwidth}
         \centering
         \includegraphics[width=\textwidth]{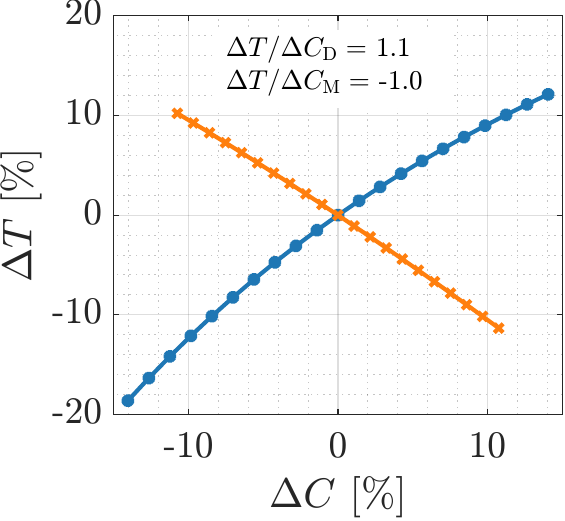}
     \end{subfigure}
          \hfill
     \begin{subfigure}{0.32\textwidth}
         \centering
         \includegraphics[width=\textwidth]{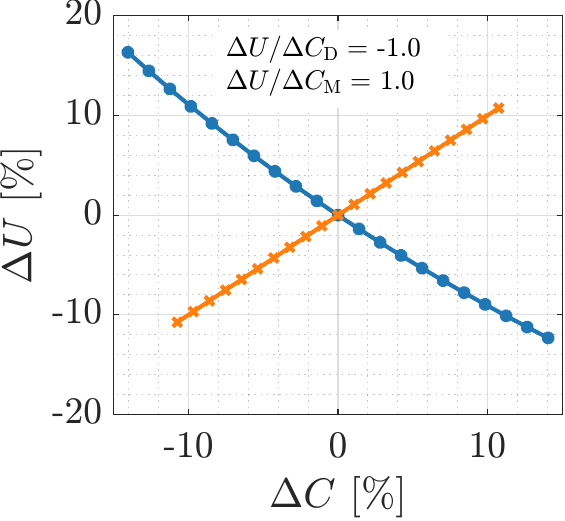}
     \end{subfigure}
    \caption{Sensitivity of $C_\mathrm{D}$ and $C_\mathrm{M}$ on nozzle pressure, temperature and velocity of the hydrogen HPDI injection at engine-relevant temperatures. The annotations provide the mean relative sensitivity coefficients.}
    \label{fig:cdcm_sensi}
\end{figure}

Figure~\ref{fig:cdcm_sensi} shows the effect of perturbations in $C_\mathrm{D}$ and $C_\mathrm{M}$ on nozzle conditions. The case at hand is the hydrogen HPDI injection at engine-relevant temperatures, and nozzle conditions are computed with the isothermal assumption. The $C_\mathrm{D}$ and $C_\mathrm{M}$ values are changed by a value in the range of $-0.1$ to $0.1$. 
The effect of changes in $C_\mathrm{D}$ and $C_\mathrm{M}$ on the nozzle conditions are opposite.
The mean relative sensitivity coefficients, which are provided in Fig.~\ref{fig:cdcm_sensi}, show that the temperature and velocity are approximately (inversely) proportional to the discharge and momentum coefficients. Actually, $\Delta U/\Delta C_\mathrm{M}$ is exactly 1, which can be derived from Eq.~\ref{eq:cdcmmach}.
Pressure scales approximately cubic with $1/C_\mathrm{D}$ and quadratic with $C_\mathrm{M}$.
It is interesting to note that similar results are obtained for the isenthalpic assumption.
To the authors' knowledge, there are no studies conducting mass or momentum flux measurements of gaseous fuels which report uncertainties. However, there are studies conducting momentum flux measurements using gasoline and diesel which report uncertainties of $\pm$\SI{10}{\percent}~\citep{cavicchi2020evaluation,postrioti2012experimental}. 
Assuming that momentum flux measurements of gaseous fuels are of similar accuracy, the uncertainty of the momentum coefficient will be \SI{10}{\percent}, which translates to an uncertainty of \SI{10}{\percent} in velocity and temperature, and \SI{20}{\percent} in pressure. This indicates that accurate momentum flux measurements are important. It also means that the accuracy of the 0D model is higher than accuracies typically encountered in measurements.
Uncertainties in mass flow rate measurements are typically much lower~\citep{peters2024characterizing}.

\section{Effect of 0D model on jet development and computational costs\label{sec:effect}}
\noindent The main motivation for the development of a 0D model is to reduce computation times of CFD simulations. However, a potential drawback of a such models is its incapability to predict spatial variations across the nozzle exit planes. In this section, we study the effects of the spatial variations and the error of the 0D model on the jet development. A simulated injection employing the 0D model is compared to an injection simulated with the internal injector geometry. They are both performed with the grid strategies described in Section~\ref{sec:computational_domains}. 

The hydrogen HPDI injection employing the multi-hole injector is chosen for this study. This setup shows significant spatial variations at the nozzle (Fig.~\ref{fig:hpdi_contours}). 
The distance a jet can freely propagate until it impinges a wall is typically 40-\SI{100}{\milli\m} in medium to heavy duty engines. For this reason, the chamber's length and diameter are increased to 100 and \SI{60}{\milli\m}, respectively.

Since the real gas properties in CFD can only handle a single species, a passive scalar is introduced in the simulation to track the injected fluid.
For this purpose, an additional transport equation is solved. The molecular diffusion coefficient of the passive scalar is calculated by the ratio of the kinematic viscosity and the molecular Schmidt number, for which a value of 0.2 is used. The sensitivity of the spatial jet development to the Schmidt number is low. The difference in jet penetration with a Schmidt number of 0.8 (a typical value for air) was observed to be negligible. 
The passive scalar is initialized with a value of 1 upstream and 0 downstream of the needle seat for the 3D injector case. In the simulation employing the GCM, the injected fluid is given a value of 1 and 0 in the vessel. 

\begin{figure}
    \centering
    \includegraphics[width=0.55\linewidth]{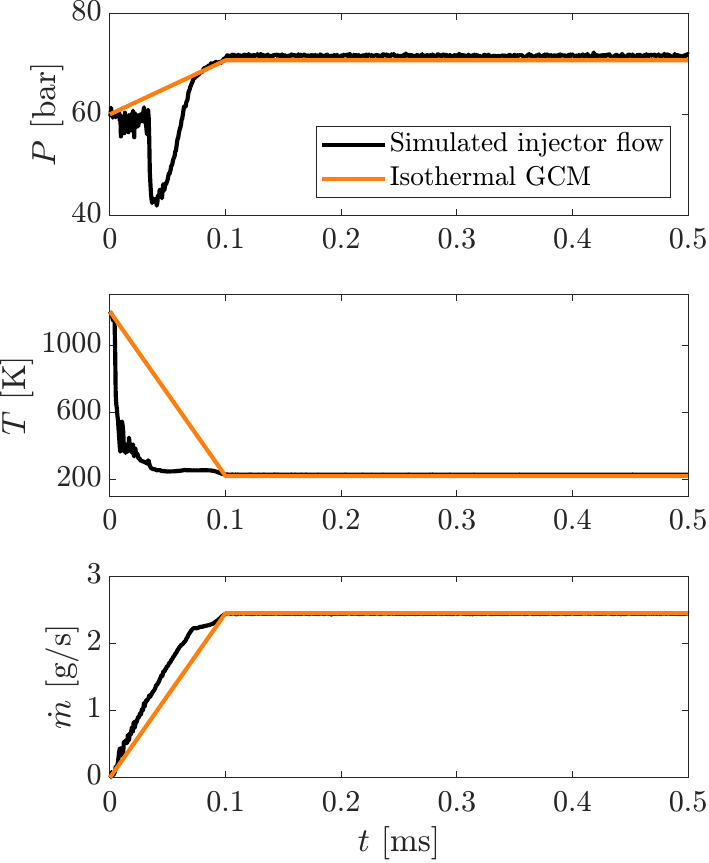}
    \caption{Simulated nozzle exit profiles and reconstructed ones using the isothermal GCM.}
    \label{fig:inflow_profs}
\end{figure}

The injection using the GCM is realized by a circular inflow boundary of \SI{0.5}{\milli\m} in diameter, which corresponds to the nozzle exit plane. Spatially uniform and time-dependent profiles of mass flow rate, pressure and temperature are imposed at this surface area. 
The profiles of the injector flow simulation are shown in Fig.~\ref{fig:inflow_profs} in black. Since the needle seat is located upstream of the nozzle exit plane, the initial nozzle exit conditions are equal to the chamber conditions. 
If the needle opening transient is mimicked by a linear process during the needle rise time (which is \SI{0.1}{\milli\s}), the inflow profiles of the simulation using the isothermal 0D model are the ones of Fig.~\ref{fig:inflow_profs} shown in orange. 

\begin{figure}
    \centering
    \includegraphics[width=0.6\linewidth]{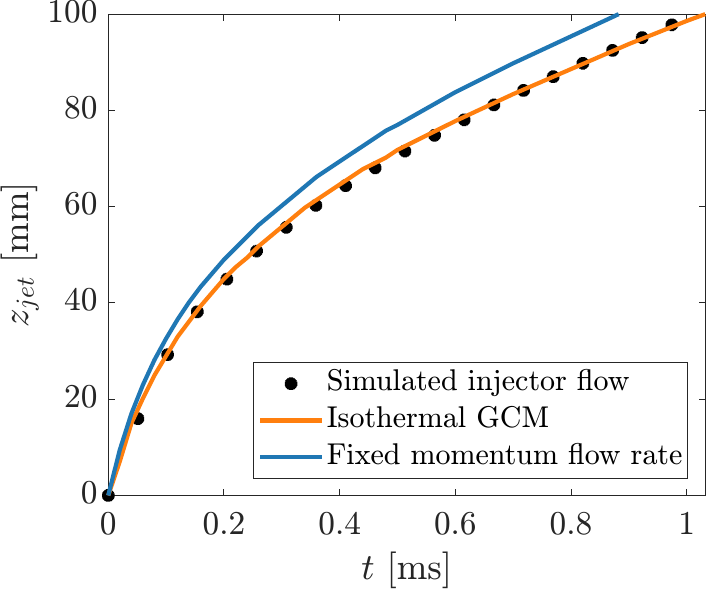}
    \caption{Jet penetration curves (denoted by $z_{jet}$) of the hydrogen HPDI injection at engine temperatures with simulated injector flow and with various nozzle exit inflow profiles.}
    \label{fig:jet_penetrations}
\end{figure}

Figure~\ref{fig:jet_penetrations} shows jet penetration data of the simulated injector flow in black and of the isothermal 0D model in orange. The outline of the jet is defined by a contour of the passive scalar with a value of 0.001. The penetration is defined as the furthest distance from the nozzle where the passive scalar is 0.001. Excellent agreement is observed between the two simulations, with a maximum error of \SI{1}{\milli\m} occurring at \SI{0.45}{\milli\s}.
As explained in Section~\ref{sec:introduction}, prior to this study, only a method existed that fixed the momentum flow rate, but not the mass flow rate~\citep{ouellette2000turbulent}. In addition, this method assumed perfect gas behavior and a sonic nozzle exit velocity, which is often not the case as is evident from Section~\ref{sec:validation} and \ref{sec:case_studies}. The jet penetration curve for this method to determine the nozzle exit conditions is plotted in Fig.~\ref{fig:jet_penetrations} in blue and shows a significant over-estimation. It shows that besides the momentum, the mass flow rate also needs to be taken into account in order to accurately predict the spatial development of the high-pressure hydrogen injection.

\begin{figure}
    \centering
    \includegraphics[width=0.5\linewidth]{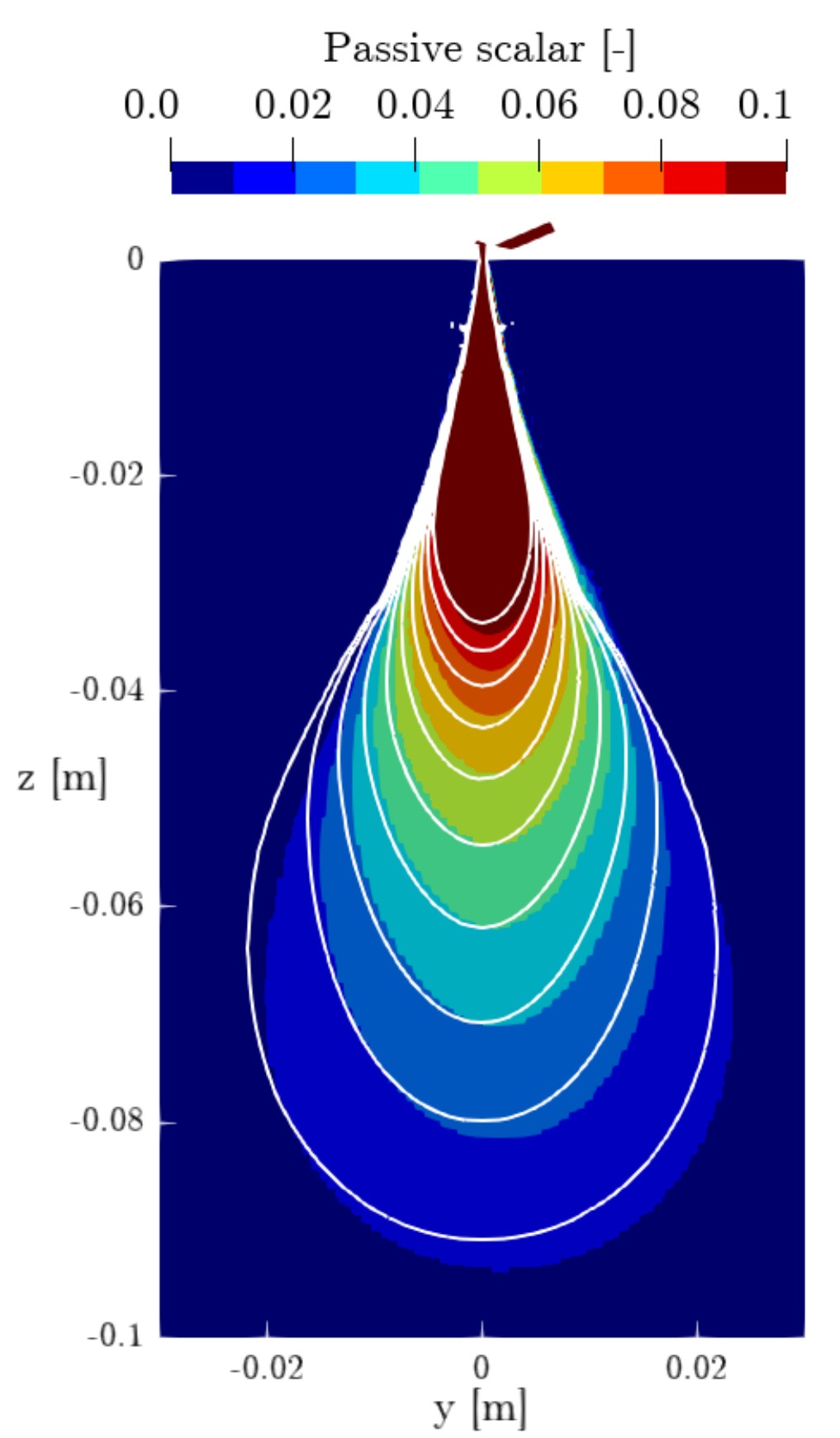}
    \caption{2D contour showing the distribution of the passive scalar of the simulated injector flow at $t=\SI{1}{\milli\s}$. The white iso-contours are the simulation result using the GCM and depict the same levels as the 2D contour. }
    \label{fig:jet_contours}
\end{figure}

Figure~\ref{fig:jet_contours} shows the spatial distribution of the passive scalar for the simulated injector flow and using the GCM for the nozzle exit conditions. Due to the spatial variations in the nozzle exit plane in the simulated injector flow (Fig.~\ref{fig:hpdi_contours}), the resulting jet is slightly asymmetric. Although the inflow of the simulation using the 0D model is uniform, the spatial distributions of the jets are in reasonable agreement. 

As mentioned at the beginning of this section, the main goal of the 0D model is to reduce computational costs associated with CFD simulations of gas injections. Table~\ref{tab:cpu_hrs} shows the costs of the simulations discussed in this section. 
It is good to note that the simulations discussed in this section were run on the same computational node which consists of 64 cores. Employing the 0D model instead of simulating the internal injector flow results in a significant reduction in computational costs. It was more than 8 times faster. 
\begin{table}
\centering
\caption{Computational costs to simulate \SI{1}{\milli\s} of hydrogen HPDI injection with internal injector flow and employing the 0D model.}
\label{tab:cpu_hrs}
\begin{tabular}{@{}lll@{}}
\toprule
Label & CPU hours [h] & \% of 3D injector \\ \midrule
3D injector   & 5,888  & 100  \\
0D model      & 704    & 12.0 \\ \bottomrule
\end{tabular}
\end{table}

\section{Conclusions}
\label{sec:conclusions}
\noindent A generic 0D global conservation model (GCM) is developed to determine nozzle exit conditions for gas injections based on reservoir conditions, generally pressure and temperature, and a discharge and momentum coefficient. 
It can be tuned for the case at hand by employing an appropriate assumption to characterize the flow upstream of the nozzle exit. 
It is validated using detailed injector flow CFD simulations with real gas properties for:
\begin{itemize}
\itemsep0em 
    \item[-] outward opening poppet-valve type, and inward opening single hole and multi hole injector,
    \item[-] injection pressures ranging from 40 to \SI{300}{\bar},
    \item[-] subsonic and choked flow injections,
    \item[-] injections in which the throat is in the nozzle or further upstream,
    \item[-] flow bench and engine conditions (elevated injector temperature).
\end{itemize}
By using the discharge and momentum coefficient as model inputs, correct determination of nozzle exit density and velocity is ensured.
Assuming that the stagnation temperature at the nozzle exit is equal to the reservoir temperature results in good agreement with the simulated nozzle exit conditions for all cases. On average, the accuracy is within \SI{1.5}{\percent}.
Also the assumption of total energy conservation between the reservoir and nozzle exit performs well.
Assuming perfect gas behavior results in larger differences, even at modest injection pressures. Use of real gas fluid properties is therefore recommended.

Mass and momentum flow rate measurements to determine the discharge and momentum coefficients are mostly conducted in a flow bench and not in the actual engine.
The CFD simulations indicate that the sensitivity of the injector wall temperature to these coefficients is low. This implicates that measurements performed at test bench conditions can be used to characterize injector flows at engine conditions.

A sensitivity analysis showed that the nozzle exit pressure changes by \SI{20}{\percent} if the momentum coefficient is disturbed by \SI{10}{\percent}. Such uncertainty is not uncommon for momentum flow measurements. For this reason, accurate values for the discharge and momentum coefficients are important.

Once an injector is characterized by experimental or numerical measurements, the model eliminates CFD simulation of internal injector flows. For a high pressure hydrogen injection, this resulted in a reduction in computation times by more than a factor 8.
Despite replacing the 3D injector by the GCM, the spatial and temporal development of the jet was simulated adequately, with negligible difference in jet penetration. 

The reduction in computation times not only allows to accelerate research and design of gas fueled internal combustion engines, it also reduces costs associated with running CFD simulations on computer facilities.
The GCM thereby aids the development of efficient and low-emission internal combustion engines.

\section*{Acknowledgements}
\noindent \noindent This publication is part of the project Argon Power Cycle (with project number 17868) of the VICI research programme which is financed by the Dutch Research Council (NWO). Computing time at the Dutch national supercomputer Snellius was funded by NWO under grant number 2024.011.

\appendix

\bibliographystyle{unsrtnat} 
\bibliography{Qiqqa2BibTexExport}

\end{document}